\documentclass[aps,pra,showpacs,twoside,twocolumn,10pt]{revtex4-2}

\usepackage[colorlinks=true, citecolor=blue, urlcolor=blue, linkcolor = blue ]{hyperref}
\usepackage{graphicx,bm,amsmath,hyperref,xcolor,braket,plain,color,amsthm,amsfonts,float,tabularx,graphicx,bbm,mathtools,esvect,wrapfig,verbatim,enumitem,dcolumn,amssymb,appendix,physics,fmtcount,booktabs,csquotes,epsfig,times,geometry,float,array,mathrsfs,graphics,epstopdf,latexsym,comment,cancel}
\usepackage[normalem]{ulem}
\usepackage[mathscr]{euscript}

\newcommand{\ga}{\gamma_h(-\omega_{h,1})}
\newcommand{\gap}{\gamma_h(\omega_{h,1})}
\newcommand{\gb}{\gamma_h(-\omega_{h,2})}
\newcommand{\gbp}{\gamma_h(\omega_{h,2})}
\newcommand{\gc}{\gamma_h(-\omega_{h,3})}
\newcommand{\gcp}{\gamma_h(\omega_{h,3})}

\newcommand{\gx}{\gamma_w(-\omega_{w,1})}
\newcommand{\gxp}{\gamma_w(\omega_{w,1})}
\newcommand{\gy}{\gamma_w(-\omega_{w,2})}
\newcommand{\gyp}{\gamma_w(\omega_{w,2})}
\newcommand{\gz}{\gamma_w(-\omega_{w,3})}
\newcommand{\gzp}{\gamma_w(\omega_{w,3})}

\newcommand{\gm}{\gamma_c(-\omega_{c,1})}
\newcommand{\gmp}{\gamma_c(\omega_{c,1})}
\newcommand{\gn}{\gamma_c(-\omega_{c,2})}
\newcommand{\gnp}{\gamma_c(\omega_{c,2})}
\newcommand{\go}{\gamma_c(-\omega_{c,3})}
\newcommand{\gop}{\gamma_c(\omega_{c,3})}

\newcommand{\Lama}{2\gbp+\gxp+2\gmp+\gyp}
\newcommand{\Lamb}{2\gb+\gnp+\gop}
\newcommand{\Lamc}{\ga+\gx+\gn+\gop}
\newcommand{\Lamd}{\gap+\gcp+2\gzp+2\gm}
\newcommand{\Lame}{\gc+\gy+\gnp+\go}
\newcommand{\Lamf}{2\gz+\gn+\go}

\newcommand{\maM}{\mathcal{M}}

\begin{document}


\title{Mpemba effect in self-contained quantum refrigerators: Accelerated cooling}

\author{Sayan Mondal}

\author{Ujjwal Sen}

\affiliation{Harish-Chandra Research Institute, A CI of Homi Bhabha National Institute, Chhatnag Road, Jhunsi, Prayagraj 211 019, India}

\begin{abstract}

 We consider the qubit-qutrit model of self-contained quantum refrigerator and observe the quantum Mpemba effect in its cooling. In this system, the qutrit acts as the refrigerator while the qubit is to be cooled. The entire system is coupled to three bosonic heat baths, due to which the dynamics of the system is governed by a Gorini-Kossakowski-Sudarshan-Lindblad master equation. We investigate the Liouvillian that generates the dynamics of the system and find that it has a block diagonal form. The dynamics of each element of the system's density matrix can be determined by solving the dynamical equation of the corresponding block that contains it. We find that the steady state belongs to the block containing only the diagonal elements in the energy basis.  We numerically solve for the steady state and investigate the steady-state cooling over a significant region of the parameter space. Moreover, we demonstrate the quantum Mpemba effect in the refrigerator: a Mpemba state obtained by applying a unitary on the equilibrium state of the system  reaches the steady state faster than the equilibrium state, despite the Mpemba state being initially farther away from the steady state. The Mpemba state thus leads to an acceleration in cooling of the cold qubit. We also find that both local and global unitaries on the qubit-qutrit system can generate the Mpemba state. Finally, we study the effect of the system-bath couplings on the Mpemba effect.     

\end{abstract} 

\maketitle

\section{Introduction}

The quantum Mpemba effect~\cite{Ares2025, Teza2025, Yu2025} is an apparently  counter-intuitive phenomenon where a state that is initially farther away from the steady state, reaches the steady state faster than some other state that was initially closer to the steady state. In the literature, the quantum Mpemba effect is usually categorized into two branches: strong Mpemba effect in open quantum systems~\cite{Carollo2021, Kochsiek2022} and symmetry restoration in closed many-body systems~\cite{Ares2023, Rylands2024}. The latter is usually observed in quantum many-body systems where the system is initialized in a symmetry-broken state and it dynamically evolves to a state with the symmetry restored. In Ref.~\cite{Ares2023}, a subsystem measure of symmetry breaking called entanglement asymmetry is introduced and it is shown that states with larger broken symmetry, restores the symmetry faster than those with lower initial symmetry breaking. This has been investigated theoretically for various models~\cite{ Yamashika2024, Chalas2024, Liu2024b, Liu2024c, Ares2025b}, as well as demonstrated experimentally~\cite{Joshi2024}. In our work we focus on the Mpemba effect in open quantum systems.    

In open quantum systems, the evolution is generated by a Liouvillian which consists of unitary part due to the system Hamiltonian as well as the dissipative part due to coupling with the environment. The eigenvalue of the Liouvillian with the largest but negative real part determines the time it takes for the system to reach the steady state, and thus the corresponding right eigenvector gives the slowest decaying mode of the Liouvillian. 
To obtain the quantum Mpemba effect, an initial state is considered with suppressed amplitude for the slowest decaying mode, so that  the time required by the state to reach steady state reduces. The effect has been studied in quantum dot systems~\cite{Chatterjee2023, Graf2025, Wang2024, Zatsarynna2025}, few-level systems~\cite{Manikandan2021, Ivander2023, Zhou2023, Chatterjee2024, Kheirandish2024, Nava2024}, quantum harmonic oscillators~\cite{Longhi2025b}, Dicke model~\cite{Carollo2021}, spin models~\cite{Nava2019,  Bao2022}, bosonic systems~\cite{Longhi2024, Westhoff2025}, Photonic systems~\cite{Longhi2024b} etc., and  has also been demonstrated in experiments~\cite{Shapira2024, Zhang2025}. In addition to these, the thermodynamics of quantum Mpemba effect~\cite{Moroder2024} and its information-geometric analysis~\cite{Bettmann2025} has be considered. Furthermore, the  effect has been investigated in  systems with non-Markovian baths~\cite{Strachan2025}, squeezed thermal baths~\cite{Furtado2024}, noisy open systems~\cite{Zhao2025}, and initial system-environment entanglement~\cite{Longhi2025}.

{
Despite these extensive studies, the occurrence of the quantum Mpemba effect in quantum thermal machines, and in particular self-contained quantum refrigerators, remains largely unexplored. Quantum refrigerators~\cite{Linden2010,  Skrzypczyk2011, Levy2012, Brunner2014, Correa2014, Brask2015, Wang2015, Mitchison2015, Mu2017, Nimmrichter2017,  Mukhopadhyay2018, Mitchison2019, Das2019,  Hewgill2020, Ghoshal2021, Ray2023, Bhattacharyya2025, Mondkar2025} are autonomous quantum thermal devices that extract heat from a system to cool it, and have been realized experimentally in nuclear spin systems~\cite{Huang2024} and trapped ions~\cite{Maslennikov2019}. These  refrigerators can be broadly classified into two categories, measurement-based and dynamical. In measurement-based refrigerators~\cite{Yan2021, Yan2022, Konar2022, Ghosh2024}, cooling is achieved by entangling the target with an auxiliary system and performing suitable measurements. In contrast, dynamical refrigerators rely solely on system–bath interactions and autonomous evolution, without any external control or measurement~\cite{Linden2010,  Skrzypczyk2011, Levy2012, Brunner2014}.
}

{
In this work, we demonstrate the quantum Mpemba effect in a self-contained quantum refrigerator model. The model consists of a qubit and a qutrit, interacting weakly with each other and each coupled to thermal baths at different temperatures. Initially, both subsystems are in local thermal equilibrium with their respective baths, forming a product initial thermal state. The refrigerator dynamics begins when the qubit–qutrit coupling is switched on. To realize the Mpemba effect, we identify the slowest decaying mode of the Liouvillian and construct a Mpemba initial state by applying a unitary transformation to the initial thermal state such that the slowest mode is suppressed. 
Interestingly, we find that even local unitary operations are sufficient to generate such Mpemba states. We further confirm that the Mpemba state lies farther from the steady state than the initial thermal state, yet relaxes faster, thereby achieving enhanced cooling of the qubit. Finally, we analyze how the effect depends on the strengths of the system–bath and qubit–qutrit couplings.

In Ref.~\cite{Carollo2021, Kochsiek2022}, the condition of observing Mpemba effect for initial pure states was discussed. Later on, in Ref.~\cite{Moroder2024}, the authors considered the Mpemba effect with mixed initial states, where the slowest mode belongs to the block with off-diagonal elements. Hence, a unitary that makes the initial state, diagonal in the energy basis
leads to the Mpemba effect. In this work, we consider a case where the slowest mode belongs to the diagonal block and we find a general condition, which when satisfied leads to an observation of the Mpemba effect.}

The remaining part of the paper is organized as follows. In Sec.~\ref{sec-prelim}, we provide a brief overview of the qubit-qutrit model of the self-contained refrigerator and the quantum Mpemba effect. 
In Sec.~\ref{sec-refri}, we study the dynamics of the refrigerator, more specifically we reduce the Liouvillian that generates the dynamics of the individual elements of the system's density matrix into block diagonal form. This greatly simplifies computation of the dynamics of the system. From this block diagonal form, we solve for the steady state of the dynamics and numerically study the cooling ability of the refrigerator across a wide parameter regime. 
In Sec.~\ref{sec-mpemba}, we compute  the eigenvalues and the corresponding eigenvectors of the Liouvillian. The eigenvalue with largest negative real part determines the time-scale to reach the steady state. We demonstrate the Mpemba effect for the system by constructing a state where this eigenvalue is suppressed and thus it reaches to the steady state faster, demonstrating accelerated cooling. Furthermore, we also study the effect of the system-bath couplings on the Mpemba effect.  
Finally, we conclude in Sec.~\ref{sec-conc}.

\begin{figure*}[t]
    \centering
    \includegraphics[width=\linewidth]{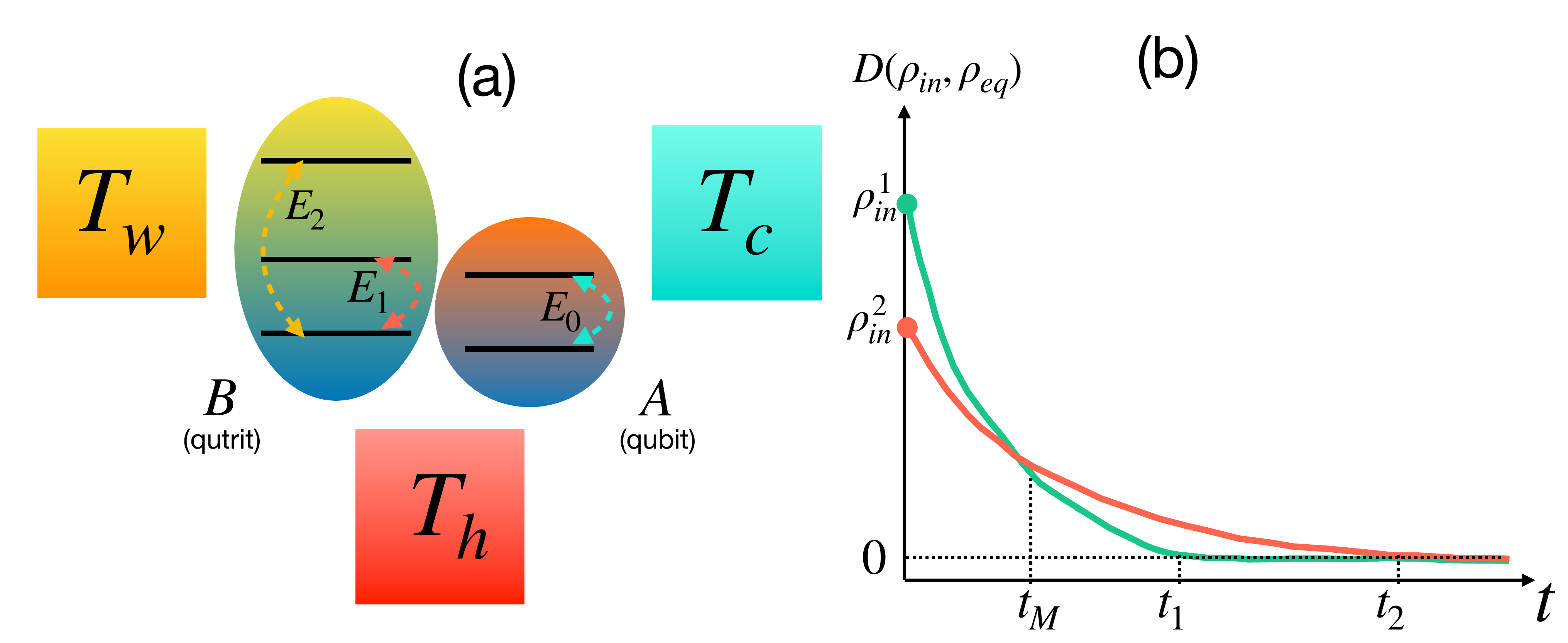}
    \caption{\textbf{\emph{Schematic diagram of self-contained refrigerator and Mpemba effect.}} {(a) We present a schematic diagram of the qubit-qutrit self-contained refrigerator coupled to three baths. The circular object labeled as $A$ is the qubit that is to be cooled, with its energy of the excited level being $E_0$. The qutrit (labeled as $B$ and act as the spiral) is given by the elliptical object with excited energy levels given by $E_1$ and $E_2$. In both qubit and qutrit, the ground state is set at zero. The three squares are the three baths, with each labeled by its temperature. The cold bath at temperature $T_c$ couples the transition between the two levels of the qubit $A$. The hot bath at $T_h$ couples the transition between the middle level and ground level of the qutrit $B$, whereas the work bath at $T_w$ couples the transition between the highest level and ground level.  
    We show the evolution of a distance-like quantity $D(\rho_{in}, \rho_{eq})$, which measures how far the system’s state is from the equilibrium steady state $\rho_{eq}$. Two initial states are considered: $\rho^{1}_{{in}}$, initially farther from equilibrium (shown in green), and $\rho^{2}_{{in}}$, initially closer to equilibrium (shown in red). Despite starting farther away, $\rho^{1}_{{in}}$ approaches the equilibrium state faster than $\rho^{2}_{{in}}$, illustrating the Mpemba effect. The trajectories intersect at the {Mpemba time} $t_M$, at which both states are equally distant from equilibrium. When $D$ reaches zero, the system has fully relaxed to the equilibrium state. The state $\rho^{1}_{{in}}$ equilibrates at time $t_1$, while $\rho^{2}_{{in}}$ equilibrates later, at time $t_2$, with $t_2 > t_1$.}}
    \label{fig:schematic}
\end{figure*}

\section{Preliminaries}
\label{sec-prelim}
The smallest self-contained refrigerator consists of a qutrit and qubit system which are coupled to three heat baths. In this system, the qutrit acts as the refrigerator and the qubit is to be cooled. The qubit is initially at equilibrium with the bath to which it is coupled. When the coupling between the qubit and qutrit is switched on, the system evolves to specific steady state, where the qubit has a lower temperature than it initially had. In this work, we accelerate this cooling using the Mpemba effect.  In this section, we introduce the qubit-qutrit refrigerator model and the quantum Mpemba effect. 
\subsection{The quantum refrigerator model}
\label{ref_intro}
The smallest self-contained model of refrigerator introduced in Ref.~\cite{Linden2010}, consists of a qutrit acting as a refrigerator and a qubit  to be cooled. The qubit $A$ and qutrit $B$ are governed by the Hamiltonians $\tilde H_A = JE_0|1\rangle_A\langle1|$ and $\tilde H_B = J(E_1|1\rangle_B\langle1| + E_2|2\rangle_B\langle2|)$ respectively. For the working of the refrigerator a self-contained condition, $E_2 = E_0 + E_1$ is imposed on the system. Due to this, there is a degeneracy between the states $|02\rangle_{AB}$ and $|11\rangle_{AB}$. 
For the refrigerator to work, the probability of $|1\rangle_A$ is to be decreased and the probability of $|0\rangle_A$ is
to be increased. This can be done, by coupling the transition  $|0\rangle_B \leftrightarrow|1\rangle_{B}$ to a hot bath at temperature $T_h$, while the transition $|0\rangle_B \leftrightarrow|2\rangle_{B}$ is coupled to a cooler work bath at temperature $T_w$. The qutrit when at equilibrium with the two baths, has probabilities $p_1^B$ and $p_2^B$ of the two levels $|1\rangle_B$ and $|2\rangle_B$ respectively. 
In order to allow transitions of the kind $|02\rangle_{AB}\leftrightarrow |11\rangle_{AB}$, it is imperative to introduce an interaction $\tilde H_{\text{int}} = Jg(|02\rangle_{AB}\langle 11|+|11\rangle_{AB}\langle 02|)$. 
For small values of $g$, the probabilities of $|11\rangle_{AB}$ and $|02\rangle_{AB}$ are very close to each other, i.e., $p_{11}^{AB} \approx p_{02}^{AB}$.  
Since, $p_1^B > p_2^B$, in order to maintain the condition $p_{11}^{AB} \approx p_{02}^{AB}$, ground-state probability of qubit increases. For judicious choice of the bath temperatures, the probabilities of the two levels of the qubit can be tuned, such that the steady-state temperature is lower than the initial temperature of the qubit. 
This refrigerator system is presented using a schematic diagram in Fig.~\ref{fig:schematic}(a).

The total Hamiltonian $\tilde H_{AB}$ acting on the whole system is,
    $\tilde{H} = \tilde{H}_A + \tilde{H}_B + \tilde{H}_\text{int}$
Here, $E_0$, $E_1$, $E_2$ and $g$ are dimensionless quantities, such that $J$ has the necessary dimension of energy. We can write the rescaled Hamiltonian, 
\begin{align}
H =  H_A + H_B + H_\text{int},
\label{tot_ham}
\end{align}
with $H_i = \tilde H_i/J$, where $i = \{A,B,\text{int} \}$.  
Due to the degeneracy between the states $|02\rangle_{AB}$ and $|11\rangle_{AB}$ and very small value of $g$, we have $p_{11}^{AB} \approx p_{02}^{AB}$ at steady state. This leads to the qubit attaining a local equilibrium state with temperature $T_v$, such that $T_v = (E_2 -E_1)/(\frac{E_2}{T_w} - \frac{E_1}{T_h})$. The qubit attains the temperature $T_v$ at steady state. If the qubit is also coupled to a bath at temperature $T_c$, then it attains a steady-state temperature in between $T_c$ and $T_v$. If the steady-state temperature is lower than the initial temperature of the qubit, then qubit is said to be cooled and the qutrit is working as a refrigerator. 
In addition to cooling qubits, the ability of self-contained refrigerators to cool higher dimensional systems has been investigated~\cite{Brunner2012, Ghanavati2014, Konar2023, Krishnan2024}. 
The discussed qutrit-qubit system has also been used for other quantum technologies like quantum transistors~\cite{Guo2018}.

As discussed earlier, the qutrit is coupled to two baths, while the qubit is coupled to one bath. The bath Hamiltonian is given by $H_{\mu} = \sum_{k} \omega_{\mu k} b^\dagger_{\mu k} b_{\mu k}$, where $\mu = \{c,h,w\}$ corresponds to the three baths $c$, $h$ and $w$ with temperatures $T_c$, $T_h$ and $T_w$ respectively. The qubit-bath and qutrit-bath interactions are given by,
\begin{align}
    H_{SB} = &\sum_k g_{ck}(b^\dagger_{ck} |0\rangle_A\langle1| + b_{ck} |1\rangle_A\langle0| )  \nonumber \\
    +&\sum_k g_{hk} (b^\dagger_{hk} |0\rangle_B\langle1| + b_{hk} |1\rangle_B\langle0| )  \nonumber \\
    +&\sum_k g_{wk} (b^\dagger_{wk} |0\rangle_B\langle2| + b_{wk} |2\rangle_B\langle0| ). 
    \label{sys-bath-hamil}
\end{align}
The total Hamiltonian of the system and bath is $H_T = \tilde H + \sum_{\mu=\{c,h,w\}} H_\mu + H_{SB}$. The dynamics of the refrigerator system is studied by employing the Born-Markov approximation. The density matrix $\rho$ of the total system $AB$ follows the Gorini-Kossakowski-Sudarshan-Lindblad (GKSL)~\cite{Petruccione2002, Nielsen2011, Rivas2012} master equation, 
$ \frac{d\rho(t)}{d\tilde t} = -\frac{i}{\hbar} [\tilde H, \rho(t)] + \sum_\mu\tilde D_\mu[\rho(t)]$ with 
$\tilde{D}_\mu[\cdot] = \sum_\omega \tilde{\gamma}_\mu(\omega) (A_{\mu,\omega} \cdot A^\dagger_{\mu,\omega} - \frac{1}{2} \{A^\dagger_{\mu,\omega}A_{\mu,\omega},\cdot\})$. Let us rescale to dimensionless quantities of  $\gamma_\mu(\omega) = \frac{\hbar}{J}\tilde \gamma_\mu(\omega)$ and $t = \frac{J}{\hbar}\tilde t$. Thus, the final form of the GKSL equation is as follows,
\begin{align}
    \frac{d\rho(t)}{dt} = -{i} [H, \rho(t)] + \sum_\mu D_\mu[\rho(t)]
    \label{eq:dyn_eq}
\end{align}
with $D_\mu[\rho] =  \sum_\omega {\gamma}_\mu(\omega) (A_{\mu,\omega} \rho A^\dagger_{\mu,\omega} - \frac{1}{2} \{A^\dagger_{\mu,\omega}A_{\mu,\omega},\rho\})$.

The decay rates are given by, 
\begin{align}
\gamma_\mu({\omega}) = 
\begin{cases}
 J_{\mu}(\omega)  \langle n_{\beta_\mu}(\omega) \rangle , & \text{if } \omega > 0, \\
 J_{\mu}(\omega) \left[\langle n_{\beta_\mu}(|\omega|) \rangle + 1\right], & \text{if } \omega < 0.
\end{cases}
\label{decay-rate}
\end{align} 
Here, we consider Ohmic baths with $J_\mu(\omega) = \kappa_\mu {|\omega|}\exp(-|\omega|/\Omega_c)$. The average photon number is given by $\langle n_{\beta_\mu}( \omega)\rangle = \frac{1}{e^{\beta_\mu\omega}-1} $, for a bath with temperature $T_\mu = 1/\beta_\mu$. The Lindblad operators $A_{\mu,\omega}$ are presented in the Appendix~\ref{app:lindbladOp}. 
\subsection{The Mpemba effect}
\label{Mpemba-intro}
In open quantum systems, the quantum Mpemba effect refers to the phenomenon where a system approaches its steady state faster when initialized in a state that is farther from the steady state than one that is closer to it. Let us consider a Markovian dynamics given by the dynamical equation,
\begin{align}
    \frac{d\rho(t)}{dt} = \mathcal{L}\rho(t).
    \label{Liouvillian}
\end{align}
Here, $\rho(t)$ is the density matrix describing the state of the system and $\mathcal{L}$ is the Liouvillian superoperator that generates the evolution the open quantum system. The solution to Eq.~\eqref{Liouvillian} is given by,
\begin{align}
    \rho(t) = e^{\mathcal{L}t}\rho(0) = \sum_{i = 1}^{d^2} \text{Tr}(l_i \rho(0))r_i e^{\lambda_i t}.
    \label{gen_sol}
\end{align}
Here, $\rho(0)$ is the initial state of the system, $\{\lambda_i\}$ are the complex eigenvalues of $\mathcal{L}$, and $\{l_i\}$ and $\{r_i\}$ are its left and right eigenstates respectively, with $\text{Tr}(l_i r_j) = \delta_{ij}$. For a $d$-dimensional system, there are $d^2$ eigenvalues and eigenstates.
When considering the Markovian assumption, $\mathcal{L}$ and consequently $\{l_i\}$, $\{r_i\}$ and $\{\lambda_i\}$ all are time-independent.  The eigenvalues are usually complex numbers with negative real numbers, except for the steady state. The steady state has eigenvalue $\lambda_1 = 0$ and the corresponding right eigenstate $r_1 = \rho_{ss}$. Moreover, the left eigenstate corresponding to the steady state is $l_1 = \mathbb{I}$, the identity operator. Thus, $\text{Tr}(l_1\rho(0)) = 1$. We can rewrite Eq.~\eqref{gen_sol} as,
\begin{align*}
    \rho(t) = \rho_{ss} + \sum_{i = 2}^{d^2} \text{Tr}(l_i\rho(0))r_i e^{\lambda_i t}.
\end{align*}
In the long time limit, $\lim_{t\rightarrow \infty} \rho(t) = \rho_{ss}$, i.e., the system reaches the steady state. The other modes have eigenvalues with real part that is negative, which leads to their amplitudes exponentially decaying with  $\lim_{t\rightarrow \infty} e^{\lambda_i t}\rightarrow 0$ for $\text{Re}[\lambda_i]<0$.

It is customary to arrange the $\lambda_i$-s in descending order of their real part, such that $0 = \text{Re}[\lambda_1] > \text{Re}[\lambda_2] \geq .... \geq \text{Re}[\lambda_{d^2}]$. The eigenvalue $\lambda_2$ has the smallest real value in magnitude, among all the non-zero eigenvalues. This is the slowest decaying mode and determines the rate at which the system reaches the steady state. More concretely, we have $|\rho(t) - \rho_{ss}| \propto \exp(\text{Re}[\lambda_2]t) = \exp(-t/\tau_2)$, with $\tau_2$ determining the decay time-scale. 
Now if we consider another initial state say $\rho_M(0) = U\rho(0)U^\dagger$, such that $\text{Tr}(l_2\rho_M(0)) = 0$. 
The evolution of this state is given by  $\rho_M(t) = e^{\mathcal{L}t}\rho_M(0)$. Since, the amplitude of $r_2$ is completely suppressed,  in this case we have  $|\rho_M(t) - \rho_{ss}| \propto \exp(\text{Re}[\lambda_3] t) = \exp(-t/\tau_3)$, with $\tau_3$ determining the decay time-scale. 
As $|\text{Re}[\lambda_2]| < |\text{Re}[\lambda_3]|$, we have $\tau_2>\tau_3$, i.e., the condition $\text{Tr}(l_2\rho_M(0)) = 0$ makes the initial state $\rho_M(0)$ reach the steady state faster than the initial state $\rho(0)$. 
In addition to this, we consider another condition that the initial state $\rho_M(0)$ is farther away from $\rho_{ss}$ when being compared with $\rho(0)$, i.e., we have $|\rho_M(0) - \rho_{ss}| > |\rho(0) - \rho_{ss}|$. The combination of the two conditions constitutes the quantum Mpemba effect. Throughout this paper, we call the initial state $\rho_M(0)$, the Mpemba state. The Mpemba effect is schematically presented in Fig.~\ref{fig:schematic}(b). We seen that there exists a specific time $t_M$ called the Mpemba time, which marks the crossing of the two trajectories. Before the Mpemba time, the distance $|\rho_M(t) - \rho_{ss}| > |\rho(t) - \rho_{ss}|$, while post Mpemba time the inequality flips.  
The quantum Mpemba effect has been observed in various quantum devices including  quantum batteries~\cite{Medina2025}, heat engines~\cite{Liu2024} and superconducting quantum computer~\cite{Edo2024}. Here we investigate the Mpemba effect in the self-contained quantum refrigerator.
 \begin{figure}[t]
    \centering
    \includegraphics[width=\linewidth]{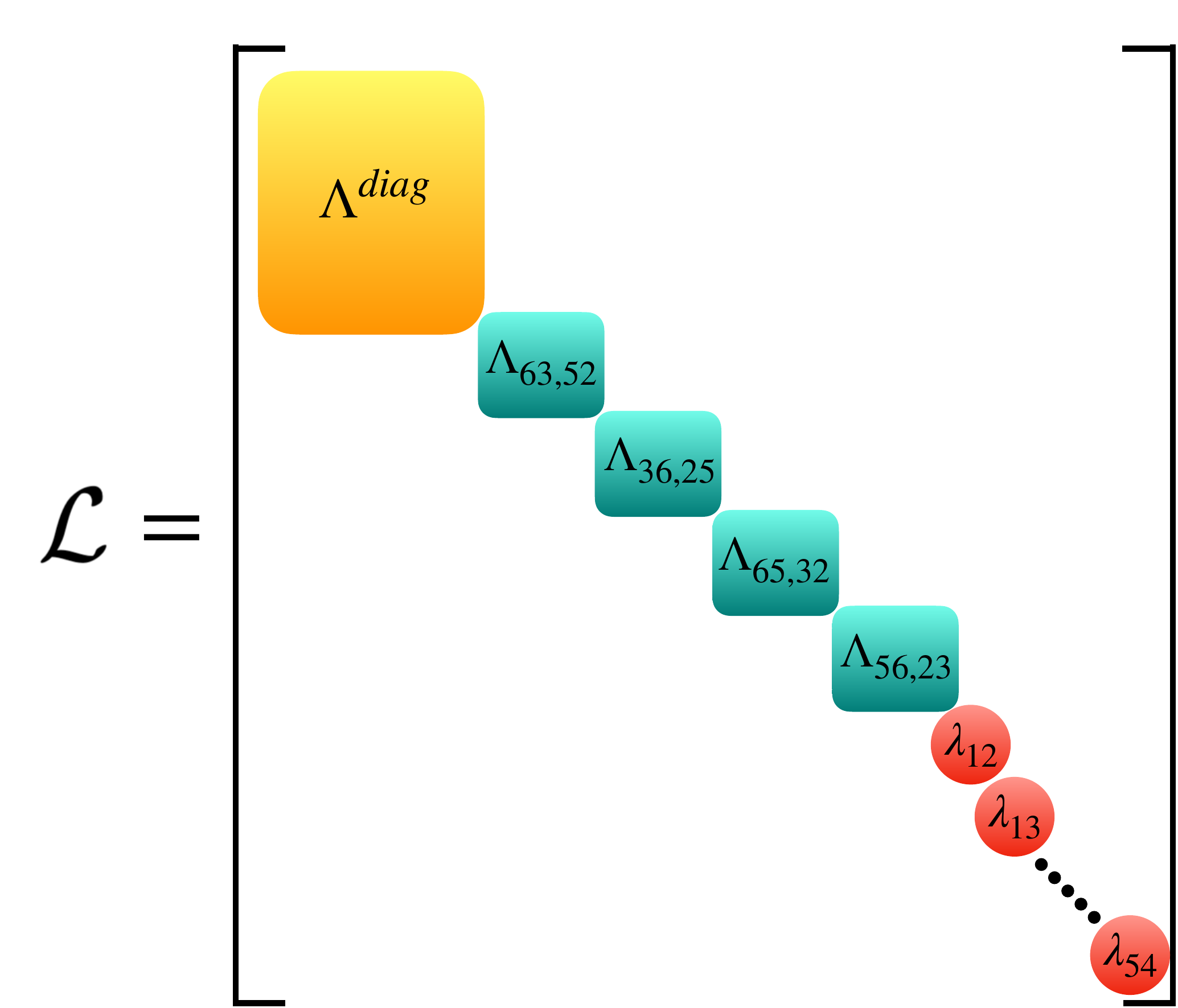}
    \caption{{\textbf{\emph{Generator of  dynamics.}} We present a schematic showing the block-diagonal form of the Liouvillian of the self-contained qubit-qutrit refrigerator. The largest block $\Lambda^\text{diag}$ in yellow, corresponds to the dynamics of the diagonal elements of the density matrix in the energy eigenbasis and is given by Eq.~\eqref{diagonal}. The smaller four blocks in green, are of dimension $2\times2$ each and are given by Eq.~\eqref{4blocks}. The other elements (shown here in red circles) do not couple to any other element and their eigenvalues $\lambda_{ij}$ are given in Eq~\eqref{liouv-eigenvals}. }}
    \label{fig:fig2a}
\end{figure} 

\begin{figure}[!h]
    \centering
    \includegraphics[width=0.6\linewidth]{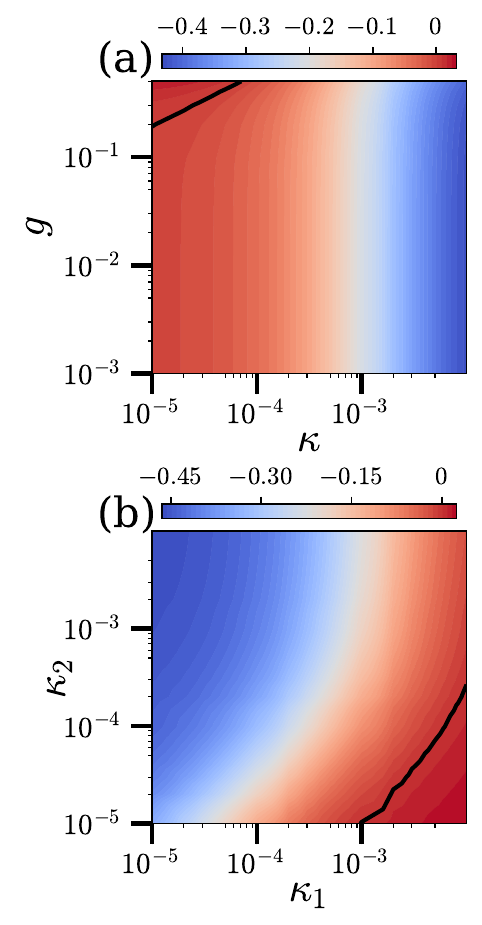}
    \caption{{\textbf{\emph{Steady state of self-contained quantum refrigerator.}}  (a) We present the change in the quantity $\Delta T = T_s - T^0_c$ with the variation in qubit-qutrit coupling $g$ and coupling between system and $h$ and $w$ bath with $\kappa_h = \kappa_w = \kappa$ and $\kappa_c = 10^{-3}$. Here, $T_s$ is the temperature of the qubit at the steady state whereas $T_c^0$ is the initial temperature of the qubit. Thus, $\Delta T$ measures the cooling capacity of the quantum refrigerator.  (b) We present the variation of $\Delta T$ with the variation of $\kappa_c = \kappa_1$ and $\kappa_h = \kappa_2$ with $\kappa_w = 10^{-3}$ and $g = 0.5$. In both the cases, $E_0 = 0.7$, $E_1 = 1.0$ with $T_h = 3.0$ and $T_c = T_w = 1.0$, consequently $T_c^0 = 1$. The solid-black line denotes $\Delta T = 0$, beyond which the system no longer provides steady-state cooling.}}
    \label{fig:steady-state}
\end{figure} 
\section{Dynamics of self-contained refrigerator}
\label{sec-refri}
In this section, we set up the dynamical equation of the elements of the density matrix of the qubit-qutrit model of the self-contained refrigerator. We observe that the dynamics is simplified due to the block diagonal form of the Liouvillian.  
Moreover, we find the steady state of the dynamics and investigate the cooling capability of the refrigerator for a significant region of the parameter space.

\subsection{Dynamical equations of self-contained refrigerator}
\label{dyn:ref}
The dynamics of the qubit-qutrit refrigerator system is governed by Eq.~\eqref{eq:dyn_eq}, where the Lindblad operators are presented in Eq.~\eqref{Lindblad-op} of the Appendix~\ref{app:lindbladOp}. We can write the state of the joint-system  as $\rho^{AB} = \sum_{i,j=1}^6\rho_{ij}|i\rangle\langle j|$, where $\{|i\rangle\}$ are the energy eigenstates as presented in Eq.~\eqref{eigenstates}.  Throughout this subsection, we consider the state to be written in the energy eigenbasis.

We investigate the action of the Liouvillian $\mathcal{L}[\cdot] = -i[H,\cdot] + \sum_\mu D_\mu[\cdot]$ on the state $\rho^{AB}$. We isolate the dynamical equation of each element $\rho_{ij}$ of the density matrix by vectorizing the density-matrix~\cite{Campaioli2024}, and obtain the corresponding matrix form of the Liouvillian. 
Thus, we have thirty-six coupled equations of the form $d\rho_{ij}/dt = [\mathcal{L}\rho]_{ij}$, where $\rho_{ij}$ are the elements of the matrix, when written in the energy eigenbasis. Depending on the form of $\mathcal{L}$, the dynamical equation of each element is determined. In the case of qubit-qutrit refrigerator system, the dynamical equations of the diagonal elements $\rho_{ii}$ decouple with the off-diagonal terms and only couple among themselves. Thus, the diagonal elements evolve independently of the off-diagonal elements. This makes the matrix $\mathcal{L}[\rho]$, block diagonal, with one of the block being $6\times6$ for the diagonal elements given by $\Lambda^\text{diag}$. The diagonal elements evolve following, $d\rho_{ii}/dt = \sum_{j = 1}^6 \Lambda^{\text{diag}}_{ij}\rho_{jj} $, with $\Lambda^{\text{diag}}$ being of the form,
\begin{widetext}
\begin{align}
\Lambda^{\text{diag}} = \frac{1}{2}
\begin{bmatrix}
-D_1 & 2\gb & \gx & 2\gm & \gy & 0 \\
2\gbp & -D_2 & \gn & 0 & \go & 0 \\
\gxp & \gnp & -D_3 & \gap & 0 & \go \\
2\gmp & 0 & \ga & -D_4 & \gc & 2\gz \\
\gyp & \gop & 0 & \gcp & -D_5 & \gn \\
0 & 0 & \gop & 2\gzp & \gnp & -D_6
\end{bmatrix},
\label{diagonal}
\end{align}
\end{widetext}
where $D_i$-s (with $i = 1,2,..,6$) are presented in Eq.~\eqref{D_values}.
The decay rates $\gamma_{\alpha}(\pm \omega_{\alpha, j})$, are defined in Eq.~\eqref{decay-rate}, with the transition energies, represented by $\omega_{c,j}$, $\omega_{h,j}$ and $\omega_{w,j}$ (with $j \in \{1,2,3\}$) defined in Eq.~\eqref{transition-energy}. On the other hand the off-diagonal terms have a much simpler dynamics. We find that certain off-diagonal terms couple with one other off-diagonal term and evolve as a pair. There are four such pairs -- $\{\rho_{52},\rho_{63}\}$, $\{\rho_{36},\rho_{25}\}$, $\{\rho_{65},\rho_{32}\}$ and $\{\rho_{56},\rho_{23}\}$. The dynamical equation of these four pairs of elements each form a $2\times2$ matrix. 
The general form of these four pairs of equation is 

\begin{align}    
    \frac{d}{dt} \begin{pmatrix}
\rho_{ij} \\
\rho_{kl}
\end{pmatrix}
&= \Lambda_{ij,kl} \begin{pmatrix}
\rho_{ij} \\
\rho_{kl}
\end{pmatrix},\quad  \text{where} \nonumber \\
    \Lambda_{63,52} &= \begin{pmatrix}
        \lambda_{63} &\gnp\\
        \gn &\lambda_{52}
    \end{pmatrix}, \nonumber \\
    \Lambda_{36,25} &= \begin{pmatrix}
        \lambda_{36} &\gnp\\
        \gn &\lambda_{25}
    \end{pmatrix}, \nonumber \\
    \Lambda_{65,32} &= \begin{pmatrix}
        \lambda_{65} &-\gop\\
        -\go &\lambda_{32}
    \end{pmatrix} \ \ \text{and}\nonumber \\
    \Lambda_{56,23} &= \begin{pmatrix}
        \lambda_{56} &-\gop\\
        -\go &\lambda_{23}
    \end{pmatrix}.
    \label{4blocks}
\end{align}
It can be seen that the matrices $\Lambda_{ij,kl}$ have the decay rates of the cold bath $c$ in its off-diagonal. So, the dynamics of these elements become completely independent when the cold bath $c$ is absent and we have $\gamma_c(\pm\omega) = 0$ for all $\omega$. 
The remaining twenty-two off-diagonal terms do not couple with each other.  The dynamical equations of these elements are of the from, $d\rho_{ij}/dt = \lambda_{ij}\rho_{ij}$ with the solution $\rho_{ij}(t) = \rho_{ij}(0)e^{\lambda_{ij}t}$, where $\lambda_{ij}$ are complex numbers with negative real parts.  The explicit form of all thirty $\lambda_{ij}$ is given by,
\begin{align}
    \lambda_{ij} &= -i\mathcal{E}_{ij} -\frac{1}{4}G_{ij}.
    \label{liouv-eigenvals}
\end{align}
Here, the first term $\mathcal{E}_{ij} = \mathcal{E}_i - \mathcal{E}_j$ is due to the unitary dynamics, whereas the second term  $G_{ij} = D_i + D_j$ is due to the coupling with the baths. The eigenvalues of the system $\mathcal{E}_i$ are given by Eq.~\eqref{hamil-eigval} and the explicit expressions of $D_i$ are given by Eq.~\eqref{D_values}. The real part of all off-diagonal elements, $\text{Re}[\lambda_{ij}] = -G_{ij}/4$, are negative.
This ensures that the off-diagonal terms decay with time and vanish at the steady state, decohering the state.  The block diagonal form of $\mathcal{L}$ is illustrated with the help of a schematic in Fig.~\ref{fig:fig2a}.

Thus, the dynamics of the thirty-six elements, reduces to solving the dynamics of one six-dimensional dynamical system, four two-dimensional system and twenty-two one-dimensional blocks. This is a much more tractable problem than a fully thirty-six-dimensional system. Although still solving the six-dimensional dynamical system analytically is very challenging. Therefore, we restore to numerical methods for solving it. We construct the left and right eigenvectors of the whole dynamics by solving for the individual block and use Eq.~\eqref{gen_sol} to given a solution for the total state $\rho^{AB}(t)$ for any time $t$. The blocks given by Eqs.~\eqref{diagonal},~\eqref{4blocks} and~\eqref{liouv-eigenvals} completely determine the dynamics of the qubit-qutrit model coupled to three heat baths. In the absence of the cold bath $c$, we can just set the decay rates corresponding to the $c$-bath to zero. 

\subsection{Steady state of self-contained refrigerator}
The qutrit-qubit refrigerator system has a unique steady state, which it attains after a long time. In this subsection, we find the steady state of the system which corresponds to the right-eigenvector of the Liouvillian $\mathcal{L}$ with zero-eigenvalue. We observe that the block-diagonal $\Lambda^\text{diag}$ corresponding to the evolution of diagonal entries $(\rho_{ii}(t))$ of the density-matrix of the system has a zero-eigenvalue. Thus the steady state $\tau$ is diagonal in the energy eigenbasis, i.e., $\tau = \sum_{i=1}^6\tau_i |i\rangle\langle i|$. We need to solve the linear equation,
\begin{align}
    \Lambda^\text{diag}\tau = 0,
    \label{steady-state}
\end{align}
with the additional constraint of unit trace, i.e., $\sum_{i = 1}^6 \tau_i = 1$.
 Although Eq.~\eqref{steady-state} is analytically solvable, the complete expression of $\tau$ is tedious and difficult to present here. Therefore, we resort to numerically solve the set of simultaneous linear equation obtained from Eq.~\eqref{steady-state} while respecting the normalization condition on trace. 
 For the qubit-qutrit system to act as a refrigerator, it is essential that the steady-state temperature $T_s$ of the qubit is lower than the initial temperature of the qubit $T_c^0$. We define the temperature of the qubit as,
 \begin{align}
    T(t) = \frac{E_0}{k_B\ln\left[p_0^A(t)/p_1^A(t)\right]},
     \label{temp-qubit}
 \end{align}
 where $k_B$ is the Boltzmann constant, $p_0^A(t)$ and $p_1^A(t)$ are the probabilities of the ground state and excited state at time $t$. Throughout this paper, we have set $k_B = 1$. Thus, the steady-state temperature and initial temperature are given by, $T_s \coloneq T(t\to\infty)$ and $T_c^0 \coloneq T(0)$ respectively. For states which are not diagonal in the energy basis, usually the definition of temperature is considered to be the temperature of the thermal state closest to the state under consideration based on the trace distance measure~\cite{Krishnan2024}. For qubits, this criteria leads to the definition in Eq.~\eqref{temp-qubit}. Thus, at least for qubits, there is no ambiguity regarding the two definitions.  
 
 We present the steady-state cooling capability of the refrigerator in  Fig.~\ref{fig:steady-state}. In our analysis, we consider the qubit-qutrit system with $E_0 = 0.7$ and $E_1 = 1.0$. Moreover, we set the temperature of the $c$, $h$ and $w$ baths to be $T_c = 1$, $T_h = 3$ and $T_w = 1$, respectively. The initial state of the qubit is taken to be the equilibrium Gibbs state at temperature $T_c$, i.e., $\rho_A(t = 0) = e^{-H_A/T_c}/\mathcal{Z}_{A,c}$, with initial temperature being $T_c^0 = T_c$. In the case, where the $c$-bath is absent and the qubit-qutrit coupling $g$ is very small, $T_s \approx T_v = (E_2-E_1)/(\frac{E_2}{T_w} -\frac{E_1}{T_h})\approx 0.5122$. 
 The cooling ability of the refrigerator on the qubit is given by $\Delta T \coloneq T_s- T_c^0 \approx -0.4878$, with the initial qubit state considered to be a Gibbs state with temperature $T_c^0 = 1$. 
 For larger values of $g$ and the presence of the $c$-bath, it is harder to estimate the steady-sate temperature of the qubit and requires solving for $\tau$ from Eq.~\eqref{steady-state}. We specifically analyze the effect of the couplings between the system and baths $\kappa_c$, $\kappa_w$ and $\kappa_h$ as well as the qubit-qutrit coupling $g$ on steady-state temperature. 
 We present the variation of $\Delta T = T_s- T_c^0$ of the qubit with system parameters in Fig.~\ref{fig:steady-state}(a) and (b). 
 In Fig.~\ref{fig:steady-state}(a), setting $\kappa_c = 10^{-3}$, we vary $\kappa_h = \kappa_w = \kappa$ and $g$, to observe that the qubit undergoes cooling except for the region of large values of $g$ and small values of $\kappa$. The contour with $\Delta T = 0$ is shown with a black solid line. 
The steady-state temperature $T_s$ is independent of $g$, when $g$ is small, while for larger values of $g$ the steady-state temperature increases with increase in $g$. On the other hand, cooling improves for larger values of $\kappa$.   
In Fig.~\ref{fig:steady-state}(b) we set $g=0.5$ and $\kappa_w = 10^{-3}$, while we vary $\kappa_c = \kappa_1$ and $\kappa_h = \kappa_2$. As in the previous case, the contour with $\Delta T = 0$ is presented in solid black line. The cooling is improved for smaller values of $\kappa_1$ and larger value of $\kappa_2$.  
For the parameter regime where $\Delta T\geq0$, the system no longer works as a refrigerator. This is consistent with the cases with smaller $g$, where the steady-state temperature decrease as the coupling with cold bath decreases and reaches its minimum the absence of the $c$-bath. 

\section{Mpemba Effect in quantum refrigerator}
\label{sec-mpemba}
In the previous section, we discussed about the dynamics generated by the Liouvillian operator and the corresponding steady state of the self-contained refrigerator. In this section, we consider the qubit-qutrit system being initialized in equilibrium with their respective baths, and investigate if it is possible to accelerate the cooling of the qubit with the help of Mpemba effect. To observe the Mpemba effect, we consider both local and global unitary operation on the initial state.  

{ 
In the qubit-qutrit refrigerator system, the slowest mode belongs to the block with diagonal elements (in the energy eigenbasis). It is quite difficult to find the algebraic form of the Mpemba unitary for such a case. Let us first give a condition that the  initial state and slowest mode should follow for observing the Mpemba effect.     
For this, we consider the dynamics starts off with the initial state $\rho_0$ and  the left eigenvector corresponding to the slowest mode is $l_2$. 
As discussed earlier, the condition for Mpemba effect is 
\begin{align}
    \text{tr}(l_2U\rho_0 {U}^\dagger) = 0.
    \label{analytic-condition1}
\end{align}
For observing the Mpemba effect, we apply the unitary $U = U_2 U_1$ on the initial state $\rho_0$, with $U_1 = \sum_i |i\rangle\langle m_i|$ and $U_2 = \sum_i |\phi_i\rangle\langle i|$. Here, $\{|i\rangle\}$ is the energy eigenstate basis of the system, with $l_2 = \sum_i \chi_i |i\rangle \langle i|$,  $\{|m_i\rangle\}$ is the eigenbasis of $\rho_0$, viz. $\rho_0 = \sum_ip_i|m_i\rangle\langle m_i|$  and $\{|\phi_i\rangle\}$ is some unknown basis that we need to find out, such that Eq.~\eqref{analytic-condition1} is true. 
The condition in Eq.~\eqref{analytic-condition1} simplifies to 
\begin{align}
    \sum_{i,j} \chi_i p_j |\langle i | \phi_j \rangle|^2 = 0.
    \label{analytic-condition2}
\end{align}
Since $\{\chi_i\}$ may have both negative and positive values while $\{p_i\}$ has only positive values, it might be possible to satisfy Eq.~\eqref{analytic-condition1}. For it to be satisfied is is necessary that,
\begin{align}
    \chi_i^\uparrow p_i^\downarrow \leq 0 \leq \chi_i^\uparrow p_i^\uparrow.
    \label{cond_mpemba}
\end{align}
Here $v^\uparrow$ and $v^\downarrow$ corresponds to arranging the elements of vector $v$ in ascending and descending order. This can be proven by first considering the fact that $M_{ij} = |\langle i|\phi_j\rangle|^2$ is a doubly stochastic matrix, as $M_{ij} \geq 0$ and  $\sum_i M_{ij} = \sum_j M_{ij} = 1$. Now, according to Birkhoff's theorem~\cite{Bhatia-book}, all doubly stochastic matrices form a convex set whose extreme points are the permutation matrices. We note that $\sum_{ij} \chi_i M_{ij} p_j$ is linear in $M_{ij}$, due to this its maximum and minimum over this convex set are attained at its extreme points. It is clear that for any permutation of the vector ${p}$, the minimum and maximum values of their inner product with vector $\chi$ is given by   $\chi_i^\uparrow p_i^\downarrow$ and $\chi_i^\uparrow p_i^\uparrow$, respectively. Hence, we get $$\chi_i^\uparrow p_i^\downarrow \leq \sum_{ij}\chi_i M_{ij} p_j \leq \chi_i^\uparrow p_i^\uparrow.$$ This along with Eq.~\eqref{analytic-condition2} leads to Eq.~\eqref{cond_mpemba}. This is a general prescription for finding the Mpemba unitary when the slowest mode and the initial state are known, and can be used beyond the system that we consider.}

The quantum refrigerator consisting of the qubit $A$ and qutrit $B$ is governed by the free Hamiltonian of $H_\text{free} = H_A + H_B$ and is coupled with three baths as discussed in Sec.~\ref{ref_intro}. Initially, the qubit-qutrit coupling is switched off. The initial state of the system is,
\begin{align}
    \rho^{AB}_\text{th}(0) = \gamma_A^{\beta_c^0} \otimes \gamma_B^{\beta_h,\beta_w},
    \label{init-state}
\end{align}
where \begin{align*}
\gamma_A^{\beta_c^0} &\coloneq \frac{1}{(1+e^{-\beta_c^0E_0})}(|0\rangle_A\langle0| + e^{-\beta_c^0E_0}|1\rangle_A\langle1|) \ \ \ \text{and} \\  
\gamma_B^{\beta_h,\beta_w} &\coloneq \frac{1}{\mathcal{Z}_{hw}}(|0\rangle_B\langle0| + e^{-\beta_hE_1}|1\rangle_B\langle1| + e^{-\beta_wE_2}|2\rangle_B\langle2|)
\end{align*}
with $\mathcal{Z}_{hw} = (1+e^{-\beta_hE_1} + e^{-\beta_wE_2})$, $\beta_\mu = 1/(k_B T_\mu)$ and $\beta_c^0 = 1/(k_B T_c(0))$, where $T_c(0)$ is the initial temperature of the cold qubit. Throughout our analysis, we set $k_B = 1$. Now at $t= 0$, the interaction $H_\text{int} = g|02\rangle_{AB}\langle11| + \text{h.c.}$, is switched on and the system evolves according to Eq.~\eqref{Liouvillian} with the total Hamiltonian being $H = H_A + H_B + H_{\text{int}}$. This leads to the system relaxing towards a specific steady state given by $\tau$ from Eq.~\eqref{steady-state}. 

On decomposing the Liouvillian $\mathcal{L}$, we observe that the initial state $\rho^{AB}_\text{th}(0)$ have contributions from the slowest eigenstate, i.e., $\text{Tr}(l_2 \rho^{AB}_\text{th}(0))\neq 0$. This means that the evolution to steady state when starting with $\rho^{AB}_\text{th}(0)$ as the initial state is dominated by $\exp[\lambda_2 t]$ term. We try to find if it is possible to start with an initial state that has no or negligible contribution from the eigenstate corresponding to $\lambda_2$ and reaches the steady state faster than $\rho^{AB}_\text{th}(0)$, leading to accelerated cooling. Moreover, for Mpemba effect another condition required is that the initial state that reaches steady state quicker also needs to be initially farther away from the steady state. For this we consider the trace-distance between the evolving state and steady state. The trace-distance measure is defined as,
\begin{align}
    D(\sigma,\eta) = \frac{1}{2}\| \sigma-\eta\|_1,
    \label{trace-distance}
\end{align}
where $\|A\|_1 = \text{Tr}(\sqrt{AA^\dagger})$. {  In Ref.~\cite{VanVu2025}, it is claimed that the Mpemba effect must be observed for all distance measures in finite time. In our case, we find that the  relative-entropy distance and trace-distance undergoes  similar time-evolution and we do observe the Mpemba effect with both, in finite time. Since, the two measures are very different in nature, we believe that the Mpemba effect will be observed in the refrigerator system for any generic distance measure.}


We demonstrate the Mpemba effect numerically in the quantum refrigerator system by comparing the dynamics of $\rho^{AB}_\text{th}(0)$ and another state  $\rho^{AB}_\text{M}(0) = U\rho^{AB}_\text{th}(0)U^\dagger$. We call this state the Mpemba state. To ensure this, we demand $D(\rho^{AB}_\text{th}(0),\tau) < D(\rho_\text{M}^{AB}(0),\tau)$ and $\rho_\text{M}^{AB}(0)$ reaches steady state faster than $\rho_\text{th}^{AB}(0)$.  

For this, we consider unitary operations that are both local and global on the qubit-qutrit system. This leads to initial states of four kinds,
\begin{subequations} \label{mpemba-initial}
\begin{align}
    \rho_{1,\text{g}}^{AB}(0) &\coloneq U_{1,\text{g}}\rho^{AB}_\text{th}(0)U_{1,\text{g}}^\dagger \label{14a} \\
    \rho_{2,\text{l}}^{AB}(0) &\coloneq U_{2,\text{l}}\rho^{AB}_\text{th}(0)U_{2,\text{l}}^\dagger \label{14b}  \\
    \rho_{3,\text{l}}^{AB}(0) &\coloneq U_{3,\text{l}}\rho^{AB}_\text{th}(0)U_{3,\text{l}}^\dagger \label{14c}  \\
    \rho_{4,\text{l}}^{AB}(0) &\coloneq U_{4,\text{l}}\rho^{AB}_\text{th}(0)U_{4,\text{l}}^\dagger. \label{14d} 
\end{align}
\end{subequations}
Here, $U_{1,\text{g}}$ is a global unitary acting on the joint qubit-qutrit system, $U_{2,\text{l}}$, $U_{3,\text{l}}$ and $U_{4,\text{l}}$ are local unitaries of the form $V_A \otimes V_B$, $V_A \otimes \mathbb{I}_B$ and $\mathbb{I}_A \otimes V_B$ respectively, with $\mathbb{I}_{A(B)}$ being the identity operator on $A(B)$ and  $V_A$ and $V_B$ being some unitary operation on $A$ and $B$ respectively. We numerically find such unitary by using Cobyla algorithm from NLOPT~\cite{NLopt}. 
We set the objective to maximize the difference between the initial distances between $D(\rho_\text{M}^{AB}(0),\tau) - D(\rho^{AB}_\text{th}(0),\tau)$, with the equality constraint that $\text{Tr}(l_2\rho_{M}^{AB}(0)) = 0$ for $\rho_M^{AB}(0)$ being the four states defined in Eq.~\eqref{mpemba-initial}. We find the optimized  unitaries for the global and local cases and their respective initial states.

{
\subsection{Mpemba effect in the absence of cold bath}
In this subsection, we consider the Mpemba effect in the absence of the $c$-bath. We analyze two cases, \emph{Case 1} - consists of additional assumptions like vanishing $g$, which makes it analytically tractable. On the other hand in \emph{Case 2}, we consider a more general setting, that can only be handled numerically.

\emph{Case 1.} Here, we assume that the qubit-qutrit coupling is very small and set $g \approx 0$. There are still too many parameters in $\Lambda_\text{diag}$, which makes it very difficult to diagonalize. Hence, in order to decrease the number of parameters, we set $E_0 = E_1 = E$, which leads to $E_2 = 2E$. In addition to this, we set $T_h = 2T_w = 2T$ and $\kappa_h = \kappa$, $\kappa_w = \kappa/2$. The qubit initially is in a local thermal state with some initial temperature $T_c(0)$. Since, the cold bath is absent, $\gamma_c(\pm \omega_{c,i}) = 0$ for all $i$. Moreover, we set we set $\Omega_c \to \infty$ in Eq.~\eqref{decay-rate}, thus $\exp(-|\omega|/\Omega_c)\approx 1$ for all $\omega$.  This eventually leads to $\gamma_h(\omega_{h,i}) = \frac{\kappa E}{e^{\beta E/2} - 1} =  \frac{\kappa E}{y - 1}$  and  $\gamma_w(\omega_{w,i}) = \frac{\kappa E}{e^{2\beta E} - 1} =  \frac{\kappa E}{y^4 - 1}$. Here we have set $y \coloneq e^{\beta E/2}$, where $\beta = 1/(k_B T)$. Substituting all these in Eq.~\eqref{diagonal}, we get 
\begin{widetext}
\begin{align}
\Lambda^{\text{diag}}_y = \frac{\kappa E}{2(y^4 - 1)}
\begin{bmatrix}
-2-2\maM & 2y\maM & y^4 & 0 & y^4 & 0 \\
2\maM & -2y\maM & 0 & 0 & 0 & 0 \\
1 & 0 & -y^4 - y\maM & \maM & 0 & 0 \\
0 & 0 & y\maM & -2-2\maM & y\maM & 2y^4 \\
1 & 0 & 0 & \maM & -y^4-y\maM & 0 \\
0 & 0 & 0 & 2 & 0 & -2y^4
\end{bmatrix}.
\label{diagonal-y}
\end{align}
\end{widetext}
To distinguish the diagonal block of the Liouvillian of the current \emph{Case 1} from the general case, we represent it as $\Lambda^\text{diag}_y$. Here, $\maM = (y^2+1)(y+1)$. Since now the matrix $\Lambda^\text{diag}_y$ depends on only one variable $y$, it is possible to diagonalize it. We find that it has the following eigenvalues.
\begin{align}
   \lambda_1^y &= 0, \quad \lambda_2^y = -\frac{1}{2}(\mathcal{C} - \mathcal{D}). \nonumber \\
   \lambda_3^y &= -2(\mathcal{A} - \mathcal{B}), \quad \lambda_4^y = -2(\mathcal{A} + \mathcal{B}) \nonumber \\
   \lambda_5^y &= -\frac{1}{2}(\mathcal{C} + \mathcal{D}), \quad \lambda_6^y = -(y\maM + y^4).
\end{align}
Here, $\mathcal{A} \coloneq \maM + y^4 $, $\mathcal{B} \coloneq \sqrt{(1+y^2+y^3)\maM + y^4}$, $\mathcal{C} \coloneq (2y+1)\maM + 3$ and $\mathcal{D} \coloneq \sqrt{(4y^5+9y^3+9y^2+16)\maM + 9y^4}$. We denote the eigenvalues as $\lambda_i^y$ to distinguish them from the eigenvalues of the general case obtained by diagonalizing the matrix $\Lambda^\text{diag}$ from Eq.~\eqref{diagonal}. 

The right eigenvector $r^y_1$ corresponding to the zero eigenvalue $\lambda^y_1$,  is the steady state. The steady-state density matrix is $\tau^y = r_1^y =  \sum_i (p_{ss}^y)_{i}|i\rangle \langle i| $, where we have,
$$p^y_{ss} = \frac{1}{\mathcal{N}}\{ y^7, y^6, y^3, y^4, y^3, 1\},$$
with $\mathcal{N} =  1+ 2y^3 +  y^4+ y^6+ y^7 $.  The steady-state temperature of the qubit is given by
$T_c^{ss} = \frac{E}{\ln({p_0^A}/{p_1^A})}$. Here, $p_0^A = (p^y_{ss})_0 + (p^y_{ss})_1 + [(p^y_{ss})_3 + (p^y_{ss})_5]/2 
= y^7 + y^6 + y^3$ following Eq.~\eqref{temp-qubit}. From this we get 
$$T_c^{ss} = E/\ln y^3 = \frac{2}{3}T.$$
This is exactly the virtual temperature for the case we are considering, as discussed in Sec.~\ref{ref_intro}. Now we are interested the slowest mode, we found that $\lambda_2^y$ corresponds to the slowest eigenmode. The corresponding left eigenvector is $l_2^y = \sum_i \chi_i^y |i^y\rangle \langle i^y|$ with
\begin{align}
\chi_i^y =  \Bigl\{&-\frac{\mathcal{D} + \maM + 2y^4-5}{4y^4\maM}, -\frac{1}{y^3}, \frac{(3y(\maM+y^3))(\maM-y^3)}{8y^3\maM}, \nonumber\\
& \frac{2y^4 - 3\maM+\mathcal{D}-1}{4y^4}, \frac{(3y(\maM+y^3))(\maM-y^3)}{8y^3\maM},1\Bigl\}.
\end{align}
Now, we need to apply an unitary $U^y$ of the initial state $\rho^y_0 = \sum_i p_i^y |c_i^y\rangle \langle c_i^y|$, in order to observe the Mpemba effect. Here, $U^y = U_2^y U_1^y$, with $U_1^y = |i^y\rangle \langle c_i^y|$. 
 To satisfy the condition in Eq.~\eqref{analytic-condition1}, we consider 
\begin{align}
    U_2^y = 
    \begin{bmatrix}
    0 & \alpha & 0 & -\sqrt{1-\alpha^2} & 0 & 0 \\
    1 & 0 & 0 & 0 & 0 & 0 \\
    0 & \sqrt{1-\alpha^2} & 0  & \alpha& 0 & 0 \\
    0 & 0 & 1 & 0 & 0 & 0 \\
    0 & 0 & 0 & 0 & 1& 0 \\
    0 & 0 & 0 & 0 & 0 & 1
    \end{bmatrix}
    \label{analytic-unitary}
\end{align}
where $$\alpha = \frac{\chi_1 p_0+\chi_2 p_1 + \chi_3 p_2 + \chi_0 p_3 + \chi_4 p_4 + \chi_5 p_5}{\chi_2 p_1 - \chi_0 p_1 + \chi_0 p_3 - \chi_2 p_3}.$$ Such a unitary will act as a Mpemba unitary when being acted on the locally-thermal  initial state of the system.

Beyond this we numerically illustrate the current case. For this we set $E = 1$ and $T = 1$ while initial state of the qubit as given in Eq.~\eqref{init-state}, with the initial temperature of cold qubit being $T_c(0) = 0.7$. This choice of initial state satisfies the condition given in Eq.~\eqref{analytic-condition2}. In Fig.~\ref{fig:analytic}, the trace distance  between the Mpemba state $\rho_M^y = U^y \rho_0^y {U^y} ^\dagger$ is presented in dashed-orange line, while that of the thermal initial state $\rho_0^y$ (of the form of Eq.~\eqref{init-state}) is in solid-blue line. We observe that the Mpemba state is initially farther away from the steady state, but reaches the steady state faster. This demonstrates the Mpemba effect.
\begin{figure}
    \centering
\includegraphics[width=0.8\linewidth]{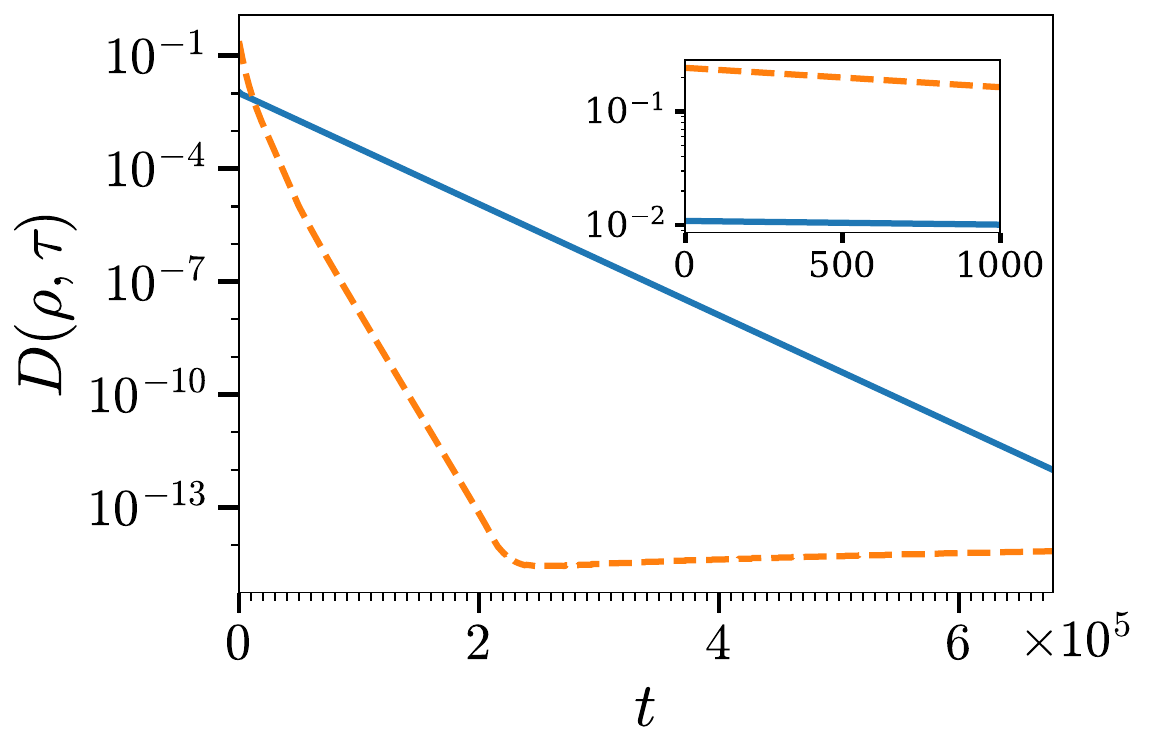}
    \caption{{\textbf{\emph{Mpemba effect in the absence of cold bath for dynamics due to $\Lambda_\text{diag}^y$.}} We consider the dynamics given by $\Lambda_y^\text{diag}$ (Eq.~\eqref{diagonal-y}), where we set $E = 1$, $T = 1$ and set the initial temperature of the cold qubit at $T_c(0) = 0.7$. The solid-blue line is the trace-distance of the state $\rho^y_0$ of the form of Eq.~\eqref{init-state}, while the dashed-orange line is the variation of the trace-distance of the Mpemba state $\rho_M^y$ from the steady state. The Mpemba state is obtained by applying the unitary $U^y = U_2^yU_1^y$ as given in Eq.~\eqref{analytic-unitary}. For the numerical demonstration we have set $g = 10^{-10}$ and $\kappa = \kappa_h = \kappa_w = 10^{-4}$. The inset magnifies the dynamics at the beginning, where we observe that the Mpemba state is initially farther away from the steady state that the thermal initial state.}}
    \label{fig:analytic}
\end{figure}

\emph{Case 2.} 
Now we consider the more general case, but still in the absence of the $c$-bath. The refrigerator is initialized in the thermal state $\tilde{\rho}_\text{th}^{AB}$ as given in Eq.~\eqref{init-state}. The Mpemba state is obtained by numerically optimizing a global unitary on qubit-qutrit system to remove the contribution of the slowest mode. The Mpemba state $\tilde\rho_{1,g}^{AB}$ is of the form of Eq.~\eqref{14a}. We numerically obtain the dynamics of the two states and compare their trace-distance from the steady state in Fig.~\ref{fig:no-bath}. For this, we set $E_0 = 0.7$, $E_1 = 1.0$, $T_c(0) = 0.7$, $T_h = 2.0$, $T_w = 1.0$. Moreover the system-bath couplings were set at $\kappa_h = \kappa_w = 10^{-4}$, while qubit-qutrit coupling was set at $g = 10^{-3}$. Consequently, every $\gamma_c(\pm \omega_{c,i}) = 0$ for all $i$. In Fig.~\ref{fig:no-bath}, the dashed-orange line corresponds to $\tilde\rho_{1,g}^{AB}$, while the solid-blue line corresponds to the $\tilde \rho_\text{th}^{AB}$. 
We observe that the distance for the Mpemba state initially decreases at a faster rate, but slows down at later times. This occurs because the slowest relaxation mode cannot be completely eliminated. Consequently, the early-time dynamics are governed by the faster mode, leading to rapid decay. However, once the fast component has decayed away, the small remaining contribution from the slowest mode takes over and determines the subsequent, slower decay rate. 
This is evident from the slope of the curves being equal at later times.
Despite this incomplete removal of the slowest mode, we still observe the Mpemba effect in this case.

}

\begin{figure}
    \centering
\includegraphics[width=0.8\linewidth]{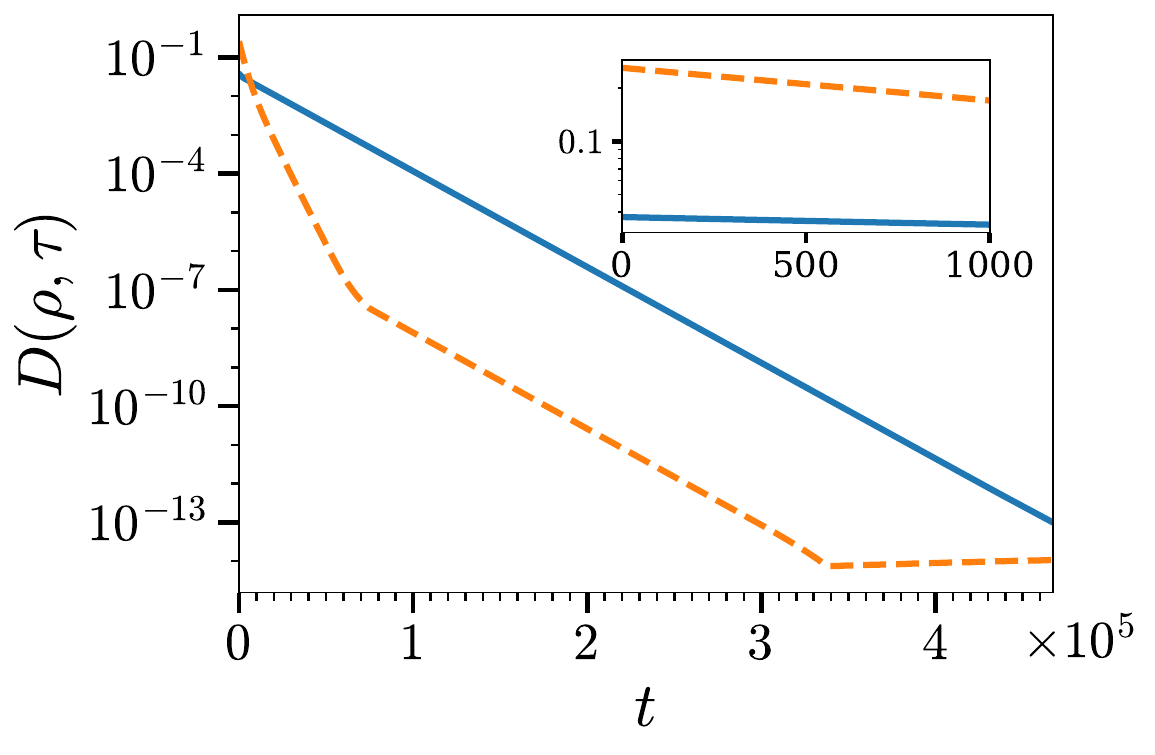}
    \caption{{\textbf{\emph{Mpemba effect in the absence of cold bath for dynamics due to the general $\Lambda_\text{diag}$. }} We present the trace distance $D(\tilde\rho_\text{th}^{AB}(t),\tau)$ in solid-blue and $D(\tilde\rho^{AB}_{1,\text{g}}(t),\tau)$ in dashed-orange line. Here $\tilde\rho_\text{th}^{AB}(t)$ and $\tilde\rho^{AB}_{1,\text{g}}(t)$ are the time-evolved states corresponding to the the initially thermal state and the Mpemba state respectively.  In the inset we clearly observe that initially, $D(\tilde\rho_\text{th}^{AB}(0),\tau) < D(\tilde\rho^{AB}_{1,\text{g}}(0),\tau)$.  The state  $\tilde\rho^{AB}_{2,\text{l}}(0)$ demonstrates the Mpemba effect and reaches the steady state faster than the initially thermal state.
    Here we have considered the dynamics due to the general $\Lambda_\text{diag}$~(Eq.~\eqref{diagonal}), with $E_0 = 0.7$, $E_1 = 1.0$, $T_c(0) = 0.7$, $T_w = 1.0$, $T_h = 2.0$, $g  =10^{-3}$, $\kappa_c = 0$ and $\kappa_h = \kappa_w = 10^{-4}$.   }}
    \label{fig:no-bath}
\end{figure}

\begin{figure}[t]
    \centering
    \includegraphics[width=0.8\linewidth]{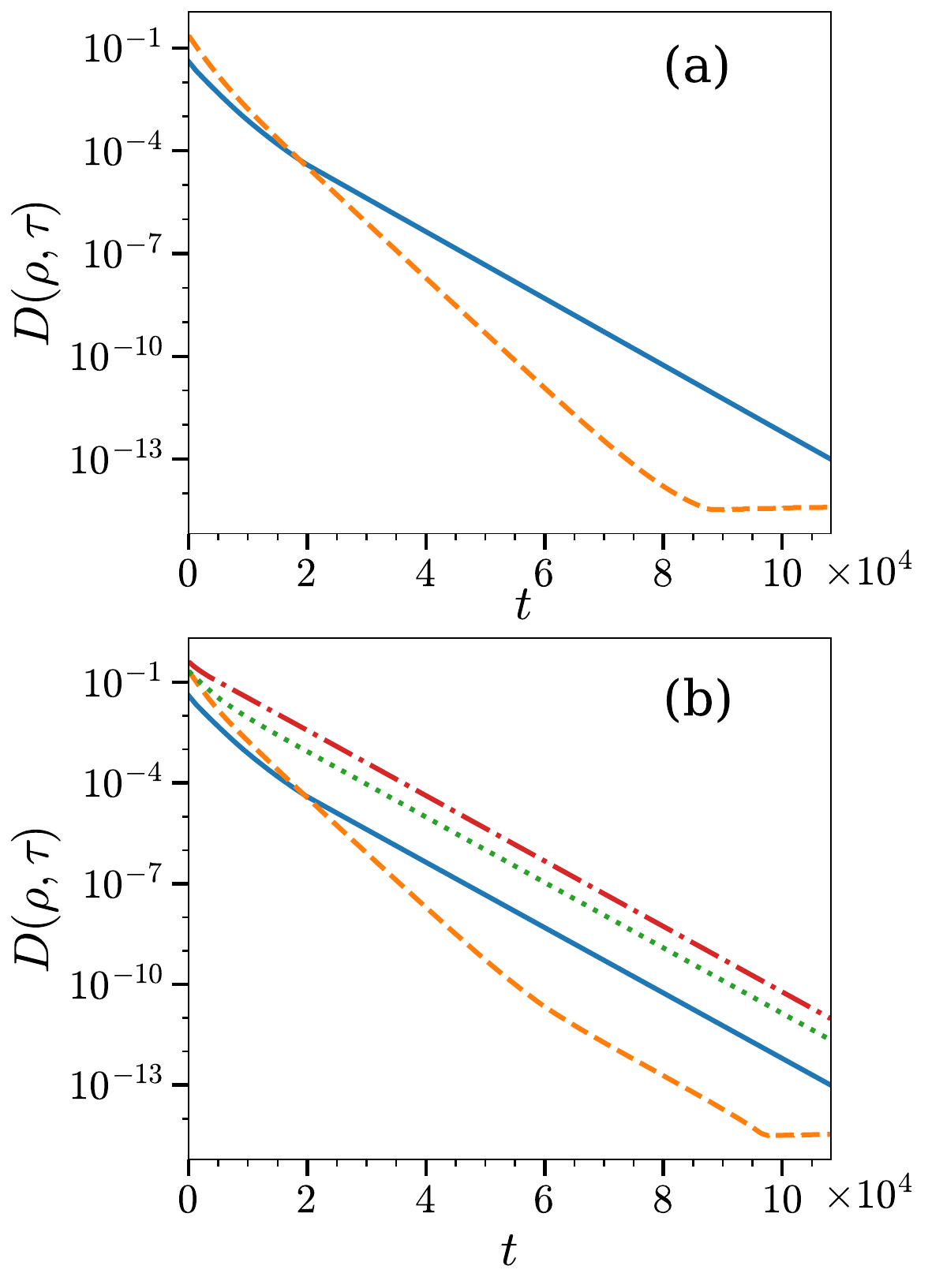}
    \caption{{\textbf{\emph{Mpemba effect in the presence of cold bath.}} (a) We present the trace distance $D(\rho_\text{th}^{AB}(t),\tau)$ in solid-blue and $D(\rho^{AB}_{1,\text{g}}(t),\tau)$ in dashed-orange line.  (b) We present the evolution of trace distance with $\tau$ of the initial states $D(\rho_\text{th}^{AB}(t),\tau)$, $D(\rho^{AB}_{2,\text{l}}(0),\tau)$, $D(\rho^{AB}_{3,\text{l}}(0),\tau)$ and $D(\rho^{AB}_{4,\text{l}}(0),\tau)$ in solid-blue, dashed-orange, dotted-green and dot-dashed-red lines respectively. The state  $\rho^{AB}_{2,\text{l}}(0)$ demonstrates the Mpemba effect, while the other two states do not show the Mpemba effect. The state $\rho^{AB}_\text{th}(0)$ is the initial thermal state given in Eq.~\eqref{init-state}, while the others are defined in Eq.~\eqref{mpemba-initial}.
    Here we have considered, $E_0 = 1.0$, $E_1 = 1.0$, $T_c = T_w = 1.0$, $T_h = 3.0$, $g  =10^{-3}$ and $\kappa_c =\kappa_h = \kappa_w = 10^{-4}$. } }
    \label{fig:mpemba}
\end{figure}

\subsection{Mpemba effect in the presence of cold bath}
In Fig.~\ref{fig:mpemba}(a) and (b) we consider different initial states and demonstrate Mpemba effect in the refrigerator system in the presence of the $c$-bath. We again demonstrate the Mpemba effect numerically. In Fig.~\ref{fig:mpemba}(a), we present the two states of the form $\rho^{AB}_\text{th}(0)$ and $\rho^{AB}_{1,\text{g}}(0)$ in solid-blue and dashed-orange line respectively. We present the trace distance of states from the steady state $\tau$ and observe that the state $\rho^{AB}_{1,\text{g}}(0)$, despite being initially farther away from the steady state, reaches steady state faster than  $\rho^{AB}_\text{th}(0)$. This is the quantum Mpemba effect. In Fig.~\ref{fig:mpemba}(b), we present the trace distance from steady state for the initial states 
$\rho_\text{th}^{AB}(0)$, $\rho^{AB}_{2,\text{l}}(0)$, $\rho^{AB}_{3,\text{l}}(0)$ and $\rho^{AB}_{4,\text{l}}(0)$ in solid-blue, dashed-orange, dotted-green and dash-dotted-red lines respectively. We observe that only  $\rho^{AB}_{2,\text{l}}(0)$ shows the Mpemba effect when compared with the initial state $\rho^{AB}_\text{th}(0)$. It was possible to suppress the slowest decaying mode with local unitary of the form of $U_{2,\text{l}}$ but not possible with local unitaries like $U_{3,\text{l}}$ and $U_{4,\text{l}}$.
On further investigation we realize that $\rho^{AB}_{1,\text{g}}(0)$ reaches steady state faster than $\rho^{AB}_{2,\text{l}}(0)$ in all parameter regime we investigated.

Throughout Fig.~\ref{fig:mpemba}, $E_0 = 1.0$, $E_1 = 1.0$, $T_c = T_w = 1.0$, $T_h = 3.0$ and $g = 10^{-3}$. The system-bath couplings are $\kappa_h = \kappa_w = \kappa_c = 10^{-4}$. The cases presented in Fig.~\ref{fig:mpemba} is just a representative case. We do observe Mpemba effect in the quantum refrigerator for a large region of the parameter space.  

\subsection{Effect of couplings on Mpemba time}
In addition to this, we investigate the effect of the system-environment coupling $\kappa_\mu$ on the Mpemba effect observed in the refrigerator system. In Mpemba effect, the time at which the initially distant state's trajectory crosses over that of the initially nearer state is known as Mpemba time. In our analysis, we consider the Mpemba time $t_M$, when the trace distance of the two states from the steady state intersect with each other. 

\begin{figure}[t]
    \centering
    \includegraphics[width=0.8\linewidth]{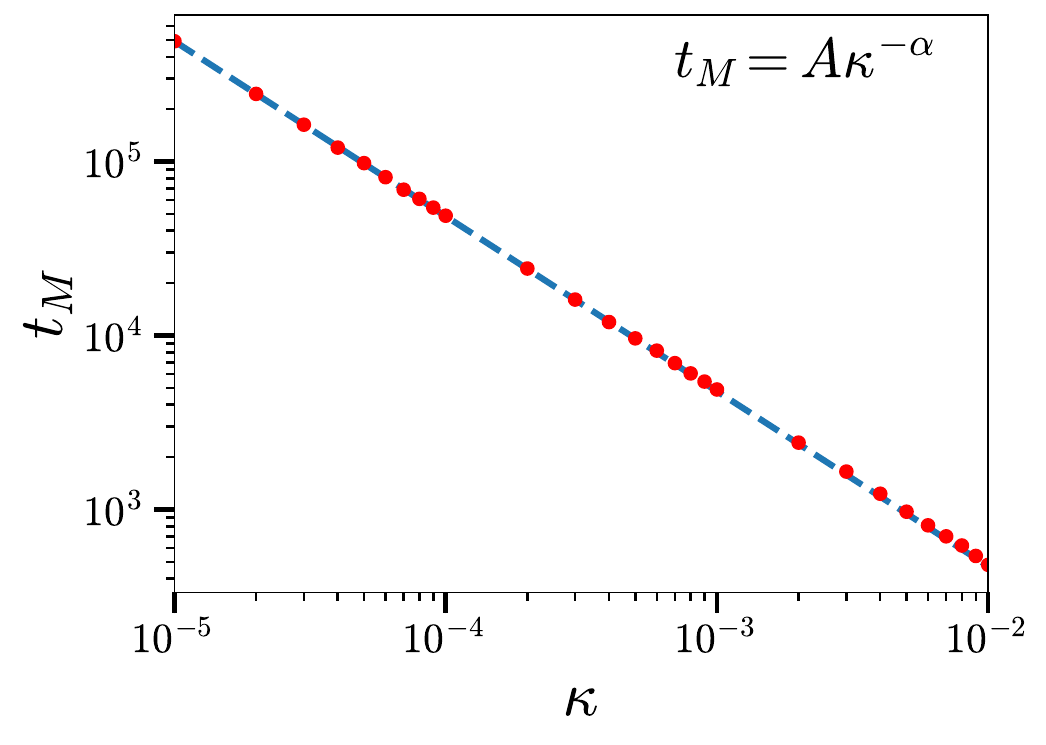}
    \caption{{\textbf{\emph{Variation of Mpemba-time in quantum refrigerator.}} The change in Mpemba-time, $t_M$ is  presented with changing system-bath couplings $\kappa$. Here we set $\kappa_c = \kappa_h = \kappa_w = \kappa$. We have set $E_0 = E_1 = 1$ with $T_c = T_w = 1$ and $T_h = 2$. The red dots are the numerically obtained data points, while solid blue line is the best fit of the form $t_M = A\kappa^{-\alpha}$. We find that $A = 4.58 \pm 0.050$ and $\alpha = 1.00 \pm 0.001$.} }
    \label{fig:mpemba-time}
\end{figure}

{In Fig.~\ref{fig:mpemba-time}, we present the variation of Mpemba time $t_M$ with the change in the system-environment couplings, all of which are set equal and varied, i.e. $\kappa_c = \kappa_w = \kappa_h = \kappa$.
We observe that $t_M$ is independent of $g$, but monotonically decreases with increase in system-environment coupling $\kappa$. The red dots corresponds to the numerically obtained values of $t_M$. We fit these values with the function, $t_M = A\kappa^{-\alpha}$. We find for the best fit, $A = 4.58 \pm 0.050$ and $\alpha = 1.00 \pm 0.001$. 
We have set the parameters as, $E_0 = E_1 = 1.0$, $T_h = 2.0$ and $T_w = T_c = 1.0$.} 
Thus, the Mpemba effect can accelerate the cooling of a qubit using quantum refrigerator.

\section{Conclusion}
\label{sec-conc}
The quantum Mpemba effect is a curious phenomenon where a state, which can be called the Mpemba state, reaches the steady state faster than some other initial state, which was initially closer to the steady state than the Mpemba state. In open quantum systems, this can be achieved by solving the dynamics of the system, specifically by finding the eigenvalues of the Liouvillian that generates the dynamics of the system. All the eigenvalues have negative real parts, and the eigenvalue with the smallest absolute value of the real part corresponds to the slowest decaying mode and determines the time it takes to reach the steady state. The Mpemba effect is realized by suppressing the amplitude of this slowest decaying mode in the initial state of the dynamics. 

In our work, we considered the qubit-qutrit self-contained quantum refrigerator and studied its dynamics. We numerically found the steady state and investigated how much cooling can be achieved for various system parameters. Subsequently, we demonstrated the Mpemba effect in this system. We found the eigenvalues and eigenvectors of the Liouvillian corresponding to the dynamics of the system. The refrigerator is usually initialized in a  state where the qubit and qutrit are locally in equilibrium with their corresponding baths.
To start the refrigeration dynamics, a qubit-qutrit coupling is switched on. We constructed both local and global unitary operators acting on this initial equilibrium state, so that the states obtained act as Mpemba states. These states are initially farther away from the steady state but reach the steady state faster than the initial equilibrium state. Thus, we can achieve the steady-state cooling quicker on starting with these Mpemba initial states. Furthermore, we studied the effect of change in system-bath couplings on the Mpemba time. 
{
Our study shows that initial thermal states  of the refrigerator system, have a non-zero projection onto the slowest Liouvillian mode. Consequently when the dynamics starts by switching on system interaction, the system relaxes slowly, even for vanishingly small interaction strength. In contrast, suitably prepared athermal states can be aligned to minimize their overlap with the slowest mode and therefore reach steady state substantially faster. This highlights that, in quantum refrigerators, steady-state time is controlled not only by thermodynamic proximity to steady-state but also by the ``distance'' of the initial state from the other dynamical eigenmodes. Thus, athermal initial state preparation can serve as a resource to accelerate cooling dynamics, constituting the physical origin and broader significance of the Mpemba effect in our model. 

This can be utilized in certain quantum devices, where reaching the steady state where, due to weak-coupling with the environment, the system takes a long time to reach the steady state. We believe that the study makes it plausible to utilize the Mpemba effect in accelerating the dynamics of such quantum devices. Thus, we can achieve the steady-state
cooling quicker on starting with these Mpemba initial states. By utilizing the Mpemba effect, such devices, like thermal machines, batteries, etc. can operate at a faster rate.  
}
\section*{Data availability}
All data used to create the figures in this
article are available in~\cite{mondal_2025_17759279}.
\appendices
\renewcommand{\theequation}{\Alph{section}\arabic{equation}}  
\setcounter{equation}{0}  
\section*{Appendices}
\section{Lindblad operators of quantum refrigerator}
\label{app:lindbladOp}
Here we present the Lindblad operators of the refrigerator system coupled with three baths.
The eigenstates of the Hamiltonian $H$ of Eq.~\eqref{tot_ham} are as follows,
\begin{align}
    |1\rangle = |00\rangle, \quad |2\rangle = |01\rangle, \quad |3\rangle = \frac{1}{\sqrt 2} (|11\rangle - |02\rangle), \nonumber\\
    |4\rangle = |10\rangle,  \quad |5\rangle = \frac{1}{\sqrt 2} (|11\rangle + |02\rangle),\quad |6\rangle = |12\rangle.
    \label{eigenstates}
\end{align}
The corresponding eigenvalues are - 
\begin{align}
\mathcal{E}_1 =  0, \quad \mathcal{E}_2 =  E_1,\quad \mathcal{E}_3 =  E_2-g ,\quad \mathcal{E}_4 =  E_0 ,\nonumber\\ \mathcal{E}_5 =  E_2+g \quad \text{and} \quad \mathcal{E}_6 =  E_0+E_2 \quad \text{respectively}.
\label{hamil-eigval}
\end{align}
For the system bath interactions given by Eq.~\eqref{sys-bath-hamil}, the GKSL equation in Eq.~\eqref{eq:dyn_eq} is derived. 
The possible transitions energies are, 
\begin{align}
    \omega_{c,1} &= E_0, \quad \omega_{c,2} = E_0-g, \quad \omega_{c,3} = E_0+g, \nonumber\\
    \omega_{h,1} &= E_1+g, \quad \omega_{h,2} = E_1, \quad \omega_{h,3} = E_1+g, \nonumber \\
    \omega_{w,1} &= E_2-g, \quad \omega_{w,2} = E_2+g, \quad \omega_{w,3} = E_2.
    \label{transition-energy}
\end{align}
The corresponding Lindblad operators are given by,
\begin{align}
    A_{c,E_1} &= |1\rangle\langle4|,\nonumber\\ 
    A_{c,E_1-g} &= \frac{1}{\sqrt 2}( |2\rangle\langle 3| +|5\rangle\langle 6|),  \nonumber\\ 
    A_{c,E_1+g} &= \frac{1}{\sqrt 2}( |2\rangle\langle 5| -|3\rangle\langle 6|) \nonumber \\
    A_{h,E_2-g} &= \frac{1}{\sqrt 2} |4\rangle\langle 3|,\quad  
    A_{h,E_2+g} = \frac{1}{\sqrt 2} |4\rangle\langle 5|\nonumber \\
     A_{h,E_2} &= |1\rangle\langle2|, \nonumber\\
    A_{w,E_3-g} &= \frac{1}{\sqrt 2} |1\rangle\langle 3|,\quad
    A_{w,E_3+g} = -\frac{1}{\sqrt 2} |1\rangle\langle 5|,\nonumber\\ 
    A_{w,E_3} &=  |4\rangle\langle 6|. 
    \label{Lindblad-op}
\end{align}
\\
\\
\section{\texorpdfstring{The expression of $D_{i}$ from $\Lambda^\text{diag}$ and $\lambda_{ij}$}{The expression of D{i}from Lambda{diag} and lambd{ij}}}
\label{lambda_ij}
Here we present the explicit expression of all $D_i$ as they appear in the expressions of  $\Lambda^\text{diag}$ and $\lambda_{ij}$ in Eq.~\eqref{diagonal} and Eq.~\eqref{liouv-eigenvals} respectively. 
The decay rates are defined in Eq.~\eqref{decay-rate} and the transition energies are defined in Eq.\eqref{transition-energy}.
\begin{align}
\label{D_values}
D_1 &= \Lama, \nonumber \\ 
D_2 &= \Lamb, \nonumber \\
D_3 &= \Lamc, \nonumber \\
D_4 &= \Lamd, \nonumber \\
D_5 &= \Lame, \text{  and}\nonumber\\
D_6 &= \Lamf. 
\end{align}

\bibliography{References}

\begin{thebibliography}{77}%
\makeatletter
\providecommand \@ifxundefined [1]{%
 \@ifx{#1\undefined}
}%
\providecommand \@ifnum [1]{%
 \ifnum #1\expandafter \@firstoftwo
 \else \expandafter \@secondoftwo
 \fi
}%
\providecommand \@ifx [1]{%
 \ifx #1\expandafter \@firstoftwo
 \else \expandafter \@secondoftwo
 \fi
}%
\providecommand \natexlab [1]{#1}%
\providecommand \enquote  [1]{``#1''}%
\providecommand \bibnamefont  [1]{#1}%
\providecommand \bibfnamefont [1]{#1}%
\providecommand \citenamefont [1]{#1}%
\providecommand \href@noop [0]{\@secondoftwo}%
\providecommand \href [0]{\begingroup \@sanitize@url \@href}%
\providecommand \@href[1]{\@@startlink{#1}\@@href}%
\providecommand \@@href[1]{\endgroup#1\@@endlink}%
\providecommand \@sanitize@url [0]{\catcode `\\12\catcode `\$12\catcode `\&12\catcode `\#12\catcode `\^12\catcode `\_12\catcode `\%12\relax}%
\providecommand \@@startlink[1]{}%
\providecommand \@@endlink[0]{}%
\providecommand \url  [0]{\begingroup\@sanitize@url \@url }%
\providecommand \@url [1]{\endgroup\@href {#1}{\urlprefix }}%
\providecommand \urlprefix  [0]{URL }%
\providecommand \Eprint [0]{\href }%
\providecommand \doibase [0]{https://doi.org/}%
\providecommand \selectlanguage [0]{\@gobble}%
\providecommand \bibinfo  [0]{\@secondoftwo}%
\providecommand \bibfield  [0]{\@secondoftwo}%
\providecommand \translation [1]{[#1]}%
\providecommand \BibitemOpen [0]{}%
\providecommand \bibitemStop [0]{}%
\providecommand \bibitemNoStop [0]{.\EOS\space}%
\providecommand \EOS [0]{\spacefactor3000\relax}%
\providecommand \BibitemShut  [1]{\csname bibitem#1\endcsname}%
\let\auto@bib@innerbib\@empty
\bibitem [{\citenamefont {Ares}\ \emph {et~al.}(2025{\natexlab{a}})\citenamefont {Ares}, \citenamefont {Calabrese},\ and\ \citenamefont {Murciano}}]{Ares2025}%
  \BibitemOpen
  \bibfield  {author} {\bibinfo {author} {\bibfnamefont {F.}~\bibnamefont {Ares}}, \bibinfo {author} {\bibfnamefont {P.}~\bibnamefont {Calabrese}},\ and\ \bibinfo {author} {\bibfnamefont {S.}~\bibnamefont {Murciano}},\ }\bibfield  {title} {\bibinfo {title} {The quantum mpemba effects},\ }\href {http://dx.doi.org/10.1038/s42254-025-00838-0} {\bibfield  {journal} {\bibinfo  {journal} {Nature Reviews Physics}\ } (\bibinfo {year} {2025}{\natexlab{a}})}\BibitemShut {NoStop}%
\bibitem [{\citenamefont {Teza}\ \emph {et~al.}(2025)\citenamefont {Teza}, \citenamefont {Bechhoefer}, \citenamefont {Lasanta}, \citenamefont {Raz},\ and\ \citenamefont {Vucelja}}]{Teza2025}%
  \BibitemOpen
  \bibfield  {author} {\bibinfo {author} {\bibfnamefont {G.}~\bibnamefont {Teza}}, \bibinfo {author} {\bibfnamefont {J.}~\bibnamefont {Bechhoefer}}, \bibinfo {author} {\bibfnamefont {A.}~\bibnamefont {Lasanta}}, \bibinfo {author} {\bibfnamefont {O.}~\bibnamefont {Raz}},\ and\ \bibinfo {author} {\bibfnamefont {M.}~\bibnamefont {Vucelja}},\ }\bibfield  {title} {\bibinfo {title} {Speedups in nonequilibrium thermal relaxation: Mpemba and related effects},\ }\href {https://doi.org/10.48550/arXiv.2502.01758} {\bibfield  {journal} {\bibinfo  {journal} {arXiv:2502.01758}\ } (\bibinfo {year} {2025})}\BibitemShut {NoStop}%
\bibitem [{\citenamefont {Yu}\ \emph {et~al.}(2025)\citenamefont {Yu}, \citenamefont {Liu},\ and\ \citenamefont {Zhang}}]{Yu2025}%
  \BibitemOpen
  \bibfield  {author} {\bibinfo {author} {\bibfnamefont {H.}~\bibnamefont {Yu}}, \bibinfo {author} {\bibfnamefont {S.}~\bibnamefont {Liu}},\ and\ \bibinfo {author} {\bibfnamefont {S.-X.}\ \bibnamefont {Zhang}},\ }\bibfield  {title} {\bibinfo {title} {Quantum mpemba effects from symmetry perspectives},\ }\bibfield  {journal} {\bibinfo  {journal} {AAPPS Bulletin}\ }\textbf {\bibinfo {volume} {35}},\ \href {https://doi.org/10.1007/s43673-025-00157-7} {10.1007/s43673-025-00157-7} (\bibinfo {year} {2025})\BibitemShut {NoStop}%
\bibitem [{\citenamefont {Carollo}\ \emph {et~al.}(2021)\citenamefont {Carollo}, \citenamefont {Lasanta},\ and\ \citenamefont {Lesanovsky}}]{Carollo2021}%
  \BibitemOpen
  \bibfield  {author} {\bibinfo {author} {\bibfnamefont {F.}~\bibnamefont {Carollo}}, \bibinfo {author} {\bibfnamefont {A.}~\bibnamefont {Lasanta}},\ and\ \bibinfo {author} {\bibfnamefont {I.}~\bibnamefont {Lesanovsky}},\ }\bibfield  {title} {\bibinfo {title} {Exponentially accelerated approach to stationarity in markovian open quantum systems through the mpemba effect},\ }\href {https://doi.org/10.1103/PhysRevLett.127.060401} {\bibfield  {journal} {\bibinfo  {journal} {Phys. Rev. Lett.}\ }\textbf {\bibinfo {volume} {127}},\ \bibinfo {pages} {060401} (\bibinfo {year} {2021})}\BibitemShut {NoStop}%
\bibitem [{\citenamefont {Kochsiek}\ \emph {et~al.}(2022)\citenamefont {Kochsiek}, \citenamefont {Carollo},\ and\ \citenamefont {Lesanovsky}}]{Kochsiek2022}%
  \BibitemOpen
  \bibfield  {author} {\bibinfo {author} {\bibfnamefont {S.}~\bibnamefont {Kochsiek}}, \bibinfo {author} {\bibfnamefont {F.}~\bibnamefont {Carollo}},\ and\ \bibinfo {author} {\bibfnamefont {I.}~\bibnamefont {Lesanovsky}},\ }\bibfield  {title} {\bibinfo {title} {Accelerating the approach of dissipative quantum spin systems towards stationarity through global spin rotations},\ }\href {https://doi.org/10.1103/PhysRevA.106.012207} {\bibfield  {journal} {\bibinfo  {journal} {Phys. Rev. A}\ }\textbf {\bibinfo {volume} {106}},\ \bibinfo {pages} {012207} (\bibinfo {year} {2022})}\BibitemShut {NoStop}%
\bibitem [{\citenamefont {Ares}\ \emph {et~al.}(2023)\citenamefont {Ares}, \citenamefont {Murciano},\ and\ \citenamefont {Calabrese}}]{Ares2023}%
  \BibitemOpen
  \bibfield  {author} {\bibinfo {author} {\bibfnamefont {F.}~\bibnamefont {Ares}}, \bibinfo {author} {\bibfnamefont {S.}~\bibnamefont {Murciano}},\ and\ \bibinfo {author} {\bibfnamefont {P.}~\bibnamefont {Calabrese}},\ }\bibfield  {title} {\bibinfo {title} {Entanglement asymmetry as a probe of symmetry breaking},\ }\href {http://dx.doi.org/10.1038/s41467-023-37747-8} {\bibfield  {journal} {\bibinfo  {journal} {Nat Commun}\ }\textbf {\bibinfo {volume} {14}} (\bibinfo {year} {2023})}\BibitemShut {NoStop}%
\bibitem [{\citenamefont {Rylands}\ \emph {et~al.}(2024)\citenamefont {Rylands}, \citenamefont {Klobas}, \citenamefont {Ares}, \citenamefont {Calabrese}, \citenamefont {Murciano},\ and\ \citenamefont {Bertini}}]{Rylands2024}%
  \BibitemOpen
  \bibfield  {author} {\bibinfo {author} {\bibfnamefont {C.}~\bibnamefont {Rylands}}, \bibinfo {author} {\bibfnamefont {K.}~\bibnamefont {Klobas}}, \bibinfo {author} {\bibfnamefont {F.}~\bibnamefont {Ares}}, \bibinfo {author} {\bibfnamefont {P.}~\bibnamefont {Calabrese}}, \bibinfo {author} {\bibfnamefont {S.}~\bibnamefont {Murciano}},\ and\ \bibinfo {author} {\bibfnamefont {B.}~\bibnamefont {Bertini}},\ }\bibfield  {title} {\bibinfo {title} {Microscopic origin of the quantum mpemba effect in integrable systems},\ }\href {https://doi.org/10.1103/PhysRevLett.133.010401} {\bibfield  {journal} {\bibinfo  {journal} {Phys. Rev. Lett.}\ }\textbf {\bibinfo {volume} {133}},\ \bibinfo {pages} {010401} (\bibinfo {year} {2024})}\BibitemShut {NoStop}%
\bibitem [{\citenamefont {Yamashika}\ \emph {et~al.}(2024)\citenamefont {Yamashika}, \citenamefont {Ares},\ and\ \citenamefont {Calabrese}}]{Yamashika2024}%
  \BibitemOpen
  \bibfield  {author} {\bibinfo {author} {\bibfnamefont {S.}~\bibnamefont {Yamashika}}, \bibinfo {author} {\bibfnamefont {F.}~\bibnamefont {Ares}},\ and\ \bibinfo {author} {\bibfnamefont {P.}~\bibnamefont {Calabrese}},\ }\bibfield  {title} {\bibinfo {title} {Entanglement asymmetry and quantum mpemba effect in two-dimensional free-fermion systems},\ }\href {https://link.aps.org/doi/10.1103/PhysRevB.110.085126} {\bibfield  {journal} {\bibinfo  {journal} {Phys. Rev. B}\ }\textbf {\bibinfo {volume} {110}},\ \bibinfo {pages} {085126} (\bibinfo {year} {2024})}\BibitemShut {NoStop}%
\bibitem [{\citenamefont {Chalas}\ \emph {et~al.}(2024)\citenamefont {Chalas}, \citenamefont {Ares}, \citenamefont {Rylands},\ and\ \citenamefont {Calabrese}}]{Chalas2024}%
  \BibitemOpen
  \bibfield  {author} {\bibinfo {author} {\bibfnamefont {K.}~\bibnamefont {Chalas}}, \bibinfo {author} {\bibfnamefont {F.}~\bibnamefont {Ares}}, \bibinfo {author} {\bibfnamefont {C.}~\bibnamefont {Rylands}},\ and\ \bibinfo {author} {\bibfnamefont {P.}~\bibnamefont {Calabrese}},\ }\bibfield  {title} {\bibinfo {title} {Multiple crossings during dynamical symmetry restoration and implications for the quantum mpemba effect},\ }\href {https://dx.doi.org/10.1088/1742-5468/ad769c} {\bibfield  {journal} {\bibinfo  {journal} {J. Stat. Mech.}\ }\textbf {\bibinfo {volume} {2024}},\ \bibinfo {pages} {103101} (\bibinfo {year} {2024})}\BibitemShut {NoStop}%
\bibitem [{\citenamefont {Liu}\ \emph {et~al.}(2024{\natexlab{a}})\citenamefont {Liu}, \citenamefont {Zhang}, \citenamefont {Yin},\ and\ \citenamefont {Zhang}}]{Liu2024b}%
  \BibitemOpen
  \bibfield  {author} {\bibinfo {author} {\bibfnamefont {S.}~\bibnamefont {Liu}}, \bibinfo {author} {\bibfnamefont {H.-K.}\ \bibnamefont {Zhang}}, \bibinfo {author} {\bibfnamefont {S.}~\bibnamefont {Yin}},\ and\ \bibinfo {author} {\bibfnamefont {S.-X.}\ \bibnamefont {Zhang}},\ }\bibfield  {title} {\bibinfo {title} {Symmetry restoration and quantum mpemba effect in symmetric random circuits},\ }\href {https://doi.org/10.1103/PhysRevLett.133.140405} {\bibfield  {journal} {\bibinfo  {journal} {Phys. Rev. Lett.}\ }\textbf {\bibinfo {volume} {133}},\ \bibinfo {pages} {140405} (\bibinfo {year} {2024}{\natexlab{a}})}\BibitemShut {NoStop}%
\bibitem [{\citenamefont {Liu}\ \emph {et~al.}(2024{\natexlab{b}})\citenamefont {Liu}, \citenamefont {Zhang}, \citenamefont {Yin}, \citenamefont {Zhang},\ and\ \citenamefont {Yao}}]{Liu2024c}%
  \BibitemOpen
  \bibfield  {author} {\bibinfo {author} {\bibfnamefont {S.}~\bibnamefont {Liu}}, \bibinfo {author} {\bibfnamefont {H.-K.}\ \bibnamefont {Zhang}}, \bibinfo {author} {\bibfnamefont {S.}~\bibnamefont {Yin}}, \bibinfo {author} {\bibfnamefont {S.-X.}\ \bibnamefont {Zhang}},\ and\ \bibinfo {author} {\bibfnamefont {H.}~\bibnamefont {Yao}},\ }\bibfield  {title} {\bibinfo {title} {Quantum mpemba effects in many-body localization systems},\ }\href {https://doi.org/10.48550/arXiv.2408.07750} {\bibfield  {journal} {\bibinfo  {journal} {arXiv:2408.07750}\ } (\bibinfo {year} {2024}{\natexlab{b}})}\BibitemShut {NoStop}%
\bibitem [{\citenamefont {Ares}\ \emph {et~al.}(2025{\natexlab{b}})\citenamefont {Ares}, \citenamefont {Vitale},\ and\ \citenamefont {Murciano}}]{Ares2025b}%
  \BibitemOpen
  \bibfield  {author} {\bibinfo {author} {\bibfnamefont {F.}~\bibnamefont {Ares}}, \bibinfo {author} {\bibfnamefont {V.}~\bibnamefont {Vitale}},\ and\ \bibinfo {author} {\bibfnamefont {S.}~\bibnamefont {Murciano}},\ }\bibfield  {title} {\bibinfo {title} {Quantum mpemba effect in free-fermionic mixed states},\ }\href {https://link.aps.org/doi/10.1103/PhysRevB.111.104312} {\bibfield  {journal} {\bibinfo  {journal} {Phys. Rev. B}\ }\textbf {\bibinfo {volume} {111}},\ \bibinfo {pages} {104312} (\bibinfo {year} {2025}{\natexlab{b}})}\BibitemShut {NoStop}%
\bibitem [{\citenamefont {Joshi}\ \emph {et~al.}(2024)\citenamefont {Joshi}, \citenamefont {Franke}, \citenamefont {Rath}, \citenamefont {Ares}, \citenamefont {Murciano}, \citenamefont {Kranzl}, \citenamefont {Blatt}, \citenamefont {Zoller}, \citenamefont {Vermersch}, \citenamefont {Calabrese}, \citenamefont {Roos},\ and\ \citenamefont {Joshi}}]{Joshi2024}%
  \BibitemOpen
  \bibfield  {author} {\bibinfo {author} {\bibfnamefont {L.~K.}\ \bibnamefont {Joshi}}, \bibinfo {author} {\bibfnamefont {J.}~\bibnamefont {Franke}}, \bibinfo {author} {\bibfnamefont {A.}~\bibnamefont {Rath}}, \bibinfo {author} {\bibfnamefont {F.}~\bibnamefont {Ares}}, \bibinfo {author} {\bibfnamefont {S.}~\bibnamefont {Murciano}}, \bibinfo {author} {\bibfnamefont {F.}~\bibnamefont {Kranzl}}, \bibinfo {author} {\bibfnamefont {R.}~\bibnamefont {Blatt}}, \bibinfo {author} {\bibfnamefont {P.}~\bibnamefont {Zoller}}, \bibinfo {author} {\bibfnamefont {B.}~\bibnamefont {Vermersch}}, \bibinfo {author} {\bibfnamefont {P.}~\bibnamefont {Calabrese}}, \bibinfo {author} {\bibfnamefont {C.~F.}\ \bibnamefont {Roos}},\ and\ \bibinfo {author} {\bibfnamefont {M.~K.}\ \bibnamefont {Joshi}},\ }\bibfield  {title} {\bibinfo {title} {Observing the quantum mpemba effect in quantum simulations},\ }\href {https://link.aps.org/doi/10.1103/PhysRevLett.133.010402} {\bibfield  {journal} {\bibinfo  {journal} {Phys. Rev. Lett.}\ }\textbf
  {\bibinfo {volume} {133}},\ \bibinfo {pages} {010402} (\bibinfo {year} {2024})}\BibitemShut {NoStop}%
\bibitem [{\citenamefont {Chatterjee}\ \emph {et~al.}(2023)\citenamefont {Chatterjee}, \citenamefont {Takada},\ and\ \citenamefont {Hayakawa}}]{Chatterjee2023}%
  \BibitemOpen
  \bibfield  {author} {\bibinfo {author} {\bibfnamefont {A.~K.}\ \bibnamefont {Chatterjee}}, \bibinfo {author} {\bibfnamefont {S.}~\bibnamefont {Takada}},\ and\ \bibinfo {author} {\bibfnamefont {H.}~\bibnamefont {Hayakawa}},\ }\bibfield  {title} {\bibinfo {title} {Quantum mpemba effect in a quantum dot with reservoirs},\ }\href {https://doi.org/10.1103/PhysRevLett.131.080402} {\bibfield  {journal} {\bibinfo  {journal} {Phys. Rev. Lett.}\ }\textbf {\bibinfo {volume} {131}},\ \bibinfo {pages} {080402} (\bibinfo {year} {2023})}\BibitemShut {NoStop}%
\bibitem [{\citenamefont {Graf}\ \emph {et~al.}(2025)\citenamefont {Graf}, \citenamefont {Splettstoesser},\ and\ \citenamefont {Monsel}}]{Graf2025}%
  \BibitemOpen
  \bibfield  {author} {\bibinfo {author} {\bibfnamefont {J.}~\bibnamefont {Graf}}, \bibinfo {author} {\bibfnamefont {J.}~\bibnamefont {Splettstoesser}},\ and\ \bibinfo {author} {\bibfnamefont {J.}~\bibnamefont {Monsel}},\ }\bibfield  {title} {\bibinfo {title} {Role of electron–electron interaction in the mpemba effect in quantum dots},\ }\href {https://doi.org/10.1088/1361-648X/adc3e3} {\bibfield  {journal} {\bibinfo  {journal} {J. Phys.: Condens. Matter}\ }\textbf {\bibinfo {volume} {37}},\ \bibinfo {pages} {195302} (\bibinfo {year} {2025})}\BibitemShut {NoStop}%
\bibitem [{\citenamefont {Wang}\ and\ \citenamefont {Wang}(2024)}]{Wang2024}%
  \BibitemOpen
  \bibfield  {author} {\bibinfo {author} {\bibfnamefont {X.}~\bibnamefont {Wang}}\ and\ \bibinfo {author} {\bibfnamefont {J.}~\bibnamefont {Wang}},\ }\bibfield  {title} {\bibinfo {title} {Mpemba effects in nonequilibrium open quantum systems},\ }\href {https://doi.org/10.1103/PhysRevResearch.6.033330} {\bibfield  {journal} {\bibinfo  {journal} {Phys. Rev. Res.}\ }\textbf {\bibinfo {volume} {6}},\ \bibinfo {pages} {033330} (\bibinfo {year} {2024})}\BibitemShut {NoStop}%
\bibitem [{\citenamefont {Zatsarynna}\ \emph {et~al.}(2025)\citenamefont {Zatsarynna}, \citenamefont {Nava}, \citenamefont {Egger},\ and\ \citenamefont {Zazunov}}]{Zatsarynna2025}%
  \BibitemOpen
  \bibfield  {author} {\bibinfo {author} {\bibfnamefont {K.}~\bibnamefont {Zatsarynna}}, \bibinfo {author} {\bibfnamefont {A.}~\bibnamefont {Nava}}, \bibinfo {author} {\bibfnamefont {R.}~\bibnamefont {Egger}},\ and\ \bibinfo {author} {\bibfnamefont {A.}~\bibnamefont {Zazunov}},\ }\bibfield  {title} {\bibinfo {title} {Green's function approach to josephson dot dynamics and application to quantum mpemba effects},\ }\href {https://doi.org/10.1103/PhysRevB.111.104506} {\bibfield  {journal} {\bibinfo  {journal} {Phys. Rev. B}\ }\textbf {\bibinfo {volume} {111}},\ \bibinfo {pages} {104506} (\bibinfo {year} {2025})}\BibitemShut {NoStop}%
\bibitem [{\citenamefont {Manikandan}(2021)}]{Manikandan2021}%
  \BibitemOpen
  \bibfield  {author} {\bibinfo {author} {\bibfnamefont {S.~K.}\ \bibnamefont {Manikandan}},\ }\bibfield  {title} {\bibinfo {title} {Equidistant quenches in few-level quantum systems},\ }\href {https://doi.org/10.1103/PhysRevResearch.3.043108} {\bibfield  {journal} {\bibinfo  {journal} {Phys. Rev. Res.}\ }\textbf {\bibinfo {volume} {3}},\ \bibinfo {pages} {043108} (\bibinfo {year} {2021})}\BibitemShut {NoStop}%
\bibitem [{\citenamefont {Ivander}\ \emph {et~al.}(2023)\citenamefont {Ivander}, \citenamefont {Anto-Sztrikacs},\ and\ \citenamefont {Segal}}]{Ivander2023}%
  \BibitemOpen
  \bibfield  {author} {\bibinfo {author} {\bibfnamefont {F.}~\bibnamefont {Ivander}}, \bibinfo {author} {\bibfnamefont {N.}~\bibnamefont {Anto-Sztrikacs}},\ and\ \bibinfo {author} {\bibfnamefont {D.}~\bibnamefont {Segal}},\ }\bibfield  {title} {\bibinfo {title} {Hyperacceleration of quantum thermalization dynamics by bypassing long-lived coherences: An analytical treatment},\ }\href {https://doi.org/10.1103/PhysRevE.108.014130} {\bibfield  {journal} {\bibinfo  {journal} {Phys. Rev. E}\ }\textbf {\bibinfo {volume} {108}},\ \bibinfo {pages} {014130} (\bibinfo {year} {2023})}\BibitemShut {NoStop}%
\bibitem [{\citenamefont {Zhou}\ \emph {et~al.}(2023)\citenamefont {Zhou}, \citenamefont {Yu}, \citenamefont {Wu}, \citenamefont {Li}, \citenamefont {Zhang}, \citenamefont {Li},\ and\ \citenamefont {Chen}}]{Zhou2023}%
  \BibitemOpen
  \bibfield  {author} {\bibinfo {author} {\bibfnamefont {Y.-L.}\ \bibnamefont {Zhou}}, \bibinfo {author} {\bibfnamefont {X.-D.}\ \bibnamefont {Yu}}, \bibinfo {author} {\bibfnamefont {C.-W.}\ \bibnamefont {Wu}}, \bibinfo {author} {\bibfnamefont {X.-Q.}\ \bibnamefont {Li}}, \bibinfo {author} {\bibfnamefont {J.}~\bibnamefont {Zhang}}, \bibinfo {author} {\bibfnamefont {W.}~\bibnamefont {Li}},\ and\ \bibinfo {author} {\bibfnamefont {P.-X.}\ \bibnamefont {Chen}},\ }\bibfield  {title} {\bibinfo {title} {Accelerating relaxation through liouvillian exceptional point},\ }\href {https://doi.org/10.1103/PhysRevResearch.5.043036} {\bibfield  {journal} {\bibinfo  {journal} {Phys. Rev. Res.}\ }\textbf {\bibinfo {volume} {5}},\ \bibinfo {pages} {043036} (\bibinfo {year} {2023})}\BibitemShut {NoStop}%
\bibitem [{\citenamefont {Chatterjee}\ \emph {et~al.}(2024)\citenamefont {Chatterjee}, \citenamefont {Takada},\ and\ \citenamefont {Hayakawa}}]{Chatterjee2024}%
  \BibitemOpen
  \bibfield  {author} {\bibinfo {author} {\bibfnamefont {A.~K.}\ \bibnamefont {Chatterjee}}, \bibinfo {author} {\bibfnamefont {S.}~\bibnamefont {Takada}},\ and\ \bibinfo {author} {\bibfnamefont {H.}~\bibnamefont {Hayakawa}},\ }\bibfield  {title} {\bibinfo {title} {Multiple quantum mpemba effect: Exceptional points and oscillations},\ }\href {https://doi.org/10.1103/PhysRevA.110.022213} {\bibfield  {journal} {\bibinfo  {journal} {Phys. Rev. A}\ }\textbf {\bibinfo {volume} {110}},\ \bibinfo {pages} {022213} (\bibinfo {year} {2024})}\BibitemShut {NoStop}%
\bibitem [{\citenamefont {Kheirandish}\ \emph {et~al.}(2024)\citenamefont {Kheirandish}, \citenamefont {Cheraghpour},\ and\ \citenamefont {Moradian}}]{Kheirandish2024}%
  \BibitemOpen
  \bibfield  {author} {\bibinfo {author} {\bibfnamefont {F.}~\bibnamefont {Kheirandish}}, \bibinfo {author} {\bibfnamefont {N.}~\bibnamefont {Cheraghpour}},\ and\ \bibinfo {author} {\bibfnamefont {A.}~\bibnamefont {Moradian}},\ }\bibfield  {title} {\bibinfo {title} {The mpemba effect in quantum oscillating and two-level systems},\ }\href {https://doi.org/10.48550/arXiv.2412.03943} {\bibfield  {journal} {\bibinfo  {journal} {arXiv:2412.03943}\ } (\bibinfo {year} {2024})}\BibitemShut {NoStop}%
\bibitem [{\citenamefont {Nava}\ and\ \citenamefont {Egger}(2024)}]{Nava2024}%
  \BibitemOpen
  \bibfield  {author} {\bibinfo {author} {\bibfnamefont {A.}~\bibnamefont {Nava}}\ and\ \bibinfo {author} {\bibfnamefont {R.}~\bibnamefont {Egger}},\ }\bibfield  {title} {\bibinfo {title} {Mpemba effects in open nonequilibrium quantum systems},\ }\href {https://doi.org/10.1103/PhysRevLett.133.136302} {\bibfield  {journal} {\bibinfo  {journal} {Phys. Rev. Lett.}\ }\textbf {\bibinfo {volume} {133}},\ \bibinfo {pages} {136302} (\bibinfo {year} {2024})}\BibitemShut {NoStop}%
\bibitem [{\citenamefont {Longhi}(2025{\natexlab{a}})}]{Longhi2025b}%
  \BibitemOpen
  \bibfield  {author} {\bibinfo {author} {\bibfnamefont {S.}~\bibnamefont {Longhi}},\ }\bibfield  {title} {\bibinfo {title} {Mpemba effect and super-accelerated thermalization in the damped quantum harmonic oscillator},\ }\href {http://dx.doi.org/10.22331/q-2025-03-26-1677} {\bibfield  {journal} {\bibinfo  {journal} {Quantum}\ }\textbf {\bibinfo {volume} {9}},\ \bibinfo {pages} {1677} (\bibinfo {year} {2025}{\natexlab{a}})}\BibitemShut {NoStop}%
\bibitem [{\citenamefont {Nava}\ and\ \citenamefont {Fabrizio}(2019)}]{Nava2019}%
  \BibitemOpen
  \bibfield  {author} {\bibinfo {author} {\bibfnamefont {A.}~\bibnamefont {Nava}}\ and\ \bibinfo {author} {\bibfnamefont {M.}~\bibnamefont {Fabrizio}},\ }\bibfield  {title} {\bibinfo {title} {Lindblad dissipative dynamics in the presence of phase coexistence},\ }\href {https://doi.org/10.1103/PhysRevB.100.125102} {\bibfield  {journal} {\bibinfo  {journal} {Phys. Rev. B}\ }\textbf {\bibinfo {volume} {100}},\ \bibinfo {pages} {125102} (\bibinfo {year} {2019})}\BibitemShut {NoStop}%
\bibitem [{\citenamefont {Bao}\ and\ \citenamefont {Hou}(2022)}]{Bao2022}%
  \BibitemOpen
  \bibfield  {author} {\bibinfo {author} {\bibfnamefont {R.}~\bibnamefont {Bao}}\ and\ \bibinfo {author} {\bibfnamefont {Z.}~\bibnamefont {Hou}},\ }\bibfield  {title} {\bibinfo {title} {Accelerating relaxation in markovian open quantum systems through quantum reset processes},\ }\href {https://doi.org/10.48550/arXiv.2212.11170} {\bibfield  {journal} {\bibinfo  {journal} {arXiv:2212.11170}\ } (\bibinfo {year} {2022})}\BibitemShut {NoStop}%
\bibitem [{\citenamefont {Longhi}(2024{\natexlab{a}})}]{Longhi2024}%
  \BibitemOpen
  \bibfield  {author} {\bibinfo {author} {\bibfnamefont {S.}~\bibnamefont {Longhi}},\ }\bibfield  {title} {\bibinfo {title} {Bosonic mpemba effect with non-classical states of light},\ }\href {http://dx.doi.org/10.1063/5.0234457} {\bibfield  {journal} {\bibinfo  {journal} {APL Quantum}\ }\textbf {\bibinfo {volume} {1}} (\bibinfo {year} {2024}{\natexlab{a}})}\BibitemShut {NoStop}%
\bibitem [{\citenamefont {Westhoff}\ \emph {et~al.}(2025)\citenamefont {Westhoff}, \citenamefont {Paeckel},\ and\ \citenamefont {Moroder}}]{Westhoff2025}%
  \BibitemOpen
  \bibfield  {author} {\bibinfo {author} {\bibfnamefont {P.}~\bibnamefont {Westhoff}}, \bibinfo {author} {\bibfnamefont {S.}~\bibnamefont {Paeckel}},\ and\ \bibinfo {author} {\bibfnamefont {M.}~\bibnamefont {Moroder}},\ }\bibfield  {title} {\bibinfo {title} {Fast and direct preparation of a genuine lattice bec via the quantum mpemba effect},\ }\href {https://doi.org/10.48550/arXiv.2504.05549} {\bibfield  {journal} {\bibinfo  {journal} {arXiv:2504.05549}\ } (\bibinfo {year} {2025})}\BibitemShut {NoStop}%
\bibitem [{\citenamefont {Longhi}(2024{\natexlab{b}})}]{Longhi2024b}%
  \BibitemOpen
  \bibfield  {author} {\bibinfo {author} {\bibfnamefont {S.}~\bibnamefont {Longhi}},\ }\bibfield  {title} {\bibinfo {title} {Photonic mpemba effect},\ }\href {https://doi.org/10.1364/ol.532503} {\bibfield  {journal} {\bibinfo  {journal} {Optics Letters}\ }\textbf {\bibinfo {volume} {49}},\ \bibinfo {pages} {5188} (\bibinfo {year} {2024}{\natexlab{b}})}\BibitemShut {NoStop}%
\bibitem [{\citenamefont {Aharony~Shapira}\ \emph {et~al.}(2024)\citenamefont {Aharony~Shapira}, \citenamefont {Shapira}, \citenamefont {Markov}, \citenamefont {Teza}, \citenamefont {Akerman}, \citenamefont {Raz},\ and\ \citenamefont {Ozeri}}]{Shapira2024}%
  \BibitemOpen
  \bibfield  {author} {\bibinfo {author} {\bibfnamefont {S.}~\bibnamefont {Aharony~Shapira}}, \bibinfo {author} {\bibfnamefont {Y.}~\bibnamefont {Shapira}}, \bibinfo {author} {\bibfnamefont {J.}~\bibnamefont {Markov}}, \bibinfo {author} {\bibfnamefont {G.}~\bibnamefont {Teza}}, \bibinfo {author} {\bibfnamefont {N.}~\bibnamefont {Akerman}}, \bibinfo {author} {\bibfnamefont {O.}~\bibnamefont {Raz}},\ and\ \bibinfo {author} {\bibfnamefont {R.}~\bibnamefont {Ozeri}},\ }\bibfield  {title} {\bibinfo {title} {Inverse mpemba effect demonstrated on a single trapped ion qubit},\ }\href {https://doi.org/10.1103/PhysRevLett.133.010403} {\bibfield  {journal} {\bibinfo  {journal} {Phys. Rev. Lett.}\ }\textbf {\bibinfo {volume} {133}},\ \bibinfo {pages} {010403} (\bibinfo {year} {2024})}\BibitemShut {NoStop}%
\bibitem [{\citenamefont {Zhang}\ \emph {et~al.}(2025)\citenamefont {Zhang}, \citenamefont {Xia}, \citenamefont {Wu}, \citenamefont {Chen}, \citenamefont {Zhang}, \citenamefont {Xie}, \citenamefont {Su}, \citenamefont {Wu}, \citenamefont {Qiu}, \citenamefont {Chen}, \citenamefont {Li}, \citenamefont {Jing},\ and\ \citenamefont {Zhou}}]{Zhang2025}%
  \BibitemOpen
  \bibfield  {author} {\bibinfo {author} {\bibfnamefont {J.}~\bibnamefont {Zhang}}, \bibinfo {author} {\bibfnamefont {G.}~\bibnamefont {Xia}}, \bibinfo {author} {\bibfnamefont {C.-W.}\ \bibnamefont {Wu}}, \bibinfo {author} {\bibfnamefont {T.}~\bibnamefont {Chen}}, \bibinfo {author} {\bibfnamefont {Q.}~\bibnamefont {Zhang}}, \bibinfo {author} {\bibfnamefont {Y.}~\bibnamefont {Xie}}, \bibinfo {author} {\bibfnamefont {W.-B.}\ \bibnamefont {Su}}, \bibinfo {author} {\bibfnamefont {W.}~\bibnamefont {Wu}}, \bibinfo {author} {\bibfnamefont {C.-W.}\ \bibnamefont {Qiu}}, \bibinfo {author} {\bibfnamefont {P.-X.}\ \bibnamefont {Chen}}, \bibinfo {author} {\bibfnamefont {W.}~\bibnamefont {Li}}, \bibinfo {author} {\bibfnamefont {H.}~\bibnamefont {Jing}},\ and\ \bibinfo {author} {\bibfnamefont {Y.-L.}\ \bibnamefont {Zhou}},\ }\bibfield  {title} {\bibinfo {title} {Observation of quantum strong mpemba effect},\ }\href {http://dx.doi.org/10.1038/s41467-024-54303-0} {\bibfield  {journal} {\bibinfo  {journal} {Nat Commun}\
  }\textbf {\bibinfo {volume} {16}} (\bibinfo {year} {2025})}\BibitemShut {NoStop}%
\bibitem [{\citenamefont {Moroder}\ \emph {et~al.}(2024)\citenamefont {Moroder}, \citenamefont {Culhane}, \citenamefont {Zawadzki},\ and\ \citenamefont {Goold}}]{Moroder2024}%
  \BibitemOpen
  \bibfield  {author} {\bibinfo {author} {\bibfnamefont {M.}~\bibnamefont {Moroder}}, \bibinfo {author} {\bibfnamefont {O.}~\bibnamefont {Culhane}}, \bibinfo {author} {\bibfnamefont {K.}~\bibnamefont {Zawadzki}},\ and\ \bibinfo {author} {\bibfnamefont {J.}~\bibnamefont {Goold}},\ }\bibfield  {title} {\bibinfo {title} {Thermodynamics of the quantum mpemba effect},\ }\href {https://doi.org/10.1103/PhysRevLett.133.140404} {\bibfield  {journal} {\bibinfo  {journal} {Phys. Rev. Lett.}\ }\textbf {\bibinfo {volume} {133}},\ \bibinfo {pages} {140404} (\bibinfo {year} {2024})}\BibitemShut {NoStop}%
\bibitem [{\citenamefont {Bettmann}\ and\ \citenamefont {Goold}(2025)}]{Bettmann2025}%
  \BibitemOpen
  \bibfield  {author} {\bibinfo {author} {\bibfnamefont {L.~P.}\ \bibnamefont {Bettmann}}\ and\ \bibinfo {author} {\bibfnamefont {J.}~\bibnamefont {Goold}},\ }\bibfield  {title} {\bibinfo {title} {Information geometry approach to quantum stochastic thermodynamics},\ }\href {https://doi.org/10.1103/PhysRevE.111.014133} {\bibfield  {journal} {\bibinfo  {journal} {Phys. Rev. E}\ }\textbf {\bibinfo {volume} {111}},\ \bibinfo {pages} {014133} (\bibinfo {year} {2025})}\BibitemShut {NoStop}%
\bibitem [{\citenamefont {Strachan}\ \emph {et~al.}(2025)\citenamefont {Strachan}, \citenamefont {Purkayastha},\ and\ \citenamefont {Clark}}]{Strachan2025}%
  \BibitemOpen
  \bibfield  {author} {\bibinfo {author} {\bibfnamefont {D.~J.}\ \bibnamefont {Strachan}}, \bibinfo {author} {\bibfnamefont {A.}~\bibnamefont {Purkayastha}},\ and\ \bibinfo {author} {\bibfnamefont {S.~R.}\ \bibnamefont {Clark}},\ }\bibfield  {title} {\bibinfo {title} {Non-markovian quantum mpemba effect},\ }\href {https://doi.org/10.1103/PhysRevLett.134.220403} {\bibfield  {journal} {\bibinfo  {journal} {Phys. Rev. Lett.}\ }\textbf {\bibinfo {volume} {134}},\ \bibinfo {pages} {220403} (\bibinfo {year} {2025})}\BibitemShut {NoStop}%
\bibitem [{\citenamefont {Furtado}\ and\ \citenamefont {Santos}(2024)}]{Furtado2024}%
  \BibitemOpen
  \bibfield  {author} {\bibinfo {author} {\bibfnamefont {J.}~\bibnamefont {Furtado}}\ and\ \bibinfo {author} {\bibfnamefont {A.~C.}\ \bibnamefont {Santos}},\ }\bibfield  {title} {\bibinfo {title} {Strong quantum mpemba effect with squeezed thermal reservoirs},\ }\href {https://doi.org/10.48550/arXiv.2411.04545} {\bibfield  {journal} {\bibinfo  {journal} {arXiv:2411.04545}\ } (\bibinfo {year} {2024})}\BibitemShut {NoStop}%
\bibitem [{\citenamefont {Zhao}\ and\ \citenamefont {Hou}(2025)}]{Zhao2025}%
  \BibitemOpen
  \bibfield  {author} {\bibinfo {author} {\bibfnamefont {M.}~\bibnamefont {Zhao}}\ and\ \bibinfo {author} {\bibfnamefont {Z.}~\bibnamefont {Hou}},\ }\bibfield  {title} {\bibinfo {title} {Noise-induced quantum mpemba effect},\ }\href {https://arxiv.org/abs/2507.11915} {\bibfield  {journal} {\bibinfo  {journal} {arXiv:2507.11915}\ } (\bibinfo {year} {2025})}\BibitemShut {NoStop}%
\bibitem [{\citenamefont {Longhi}(2025{\natexlab{b}})}]{Longhi2025}%
  \BibitemOpen
  \bibfield  {author} {\bibinfo {author} {\bibfnamefont {S.}~\bibnamefont {Longhi}},\ }\bibfield  {title} {\bibinfo {title} {Quantum mpemba effect from initial system–reservoir entanglement},\ }\href {http://dx.doi.org/10.1063/5.0266143} {\bibfield  {journal} {\bibinfo  {journal} {APL Quantum}\ }\textbf {\bibinfo {volume} {2}} (\bibinfo {year} {2025}{\natexlab{b}})}\BibitemShut {NoStop}%
\bibitem [{\citenamefont {Linden}\ \emph {et~al.}(2010)\citenamefont {Linden}, \citenamefont {Popescu},\ and\ \citenamefont {Skrzypczyk}}]{Linden2010}%
  \BibitemOpen
  \bibfield  {author} {\bibinfo {author} {\bibfnamefont {N.}~\bibnamefont {Linden}}, \bibinfo {author} {\bibfnamefont {S.}~\bibnamefont {Popescu}},\ and\ \bibinfo {author} {\bibfnamefont {P.}~\bibnamefont {Skrzypczyk}},\ }\bibfield  {title} {\bibinfo {title} {How small can thermal machines be? the smallest possible refrigerator},\ }\href {https://link.aps.org/doi/10.1103/PhysRevLett.105.130401} {\bibfield  {journal} {\bibinfo  {journal} {Phys. Rev. Lett.}\ }\textbf {\bibinfo {volume} {105}},\ \bibinfo {pages} {130401} (\bibinfo {year} {2010})}\BibitemShut {NoStop}%
\bibitem [{\citenamefont {Skrzypczyk}\ \emph {et~al.}(2011)\citenamefont {Skrzypczyk}, \citenamefont {Brunner}, \citenamefont {Linden},\ and\ \citenamefont {Popescu}}]{Skrzypczyk2011}%
  \BibitemOpen
  \bibfield  {author} {\bibinfo {author} {\bibfnamefont {P.}~\bibnamefont {Skrzypczyk}}, \bibinfo {author} {\bibfnamefont {N.}~\bibnamefont {Brunner}}, \bibinfo {author} {\bibfnamefont {N.}~\bibnamefont {Linden}},\ and\ \bibinfo {author} {\bibfnamefont {S.}~\bibnamefont {Popescu}},\ }\bibfield  {title} {\bibinfo {title} {The smallest refrigerators can reach maximal efficiency},\ }\href {https://doi.org/10.1088/1751-8113/44/49/492002} {\bibfield  {journal} {\bibinfo  {journal} {J. Phys. A: Math. Theor}\ }\textbf {\bibinfo {volume} {44}},\ \bibinfo {pages} {492002} (\bibinfo {year} {2011})}\BibitemShut {NoStop}%
\bibitem [{\citenamefont {Levy}\ and\ \citenamefont {Kosloff}(2012)}]{Levy2012}%
  \BibitemOpen
  \bibfield  {author} {\bibinfo {author} {\bibfnamefont {A.}~\bibnamefont {Levy}}\ and\ \bibinfo {author} {\bibfnamefont {R.}~\bibnamefont {Kosloff}},\ }\bibfield  {title} {\bibinfo {title} {Quantum absorption refrigerator},\ }\href {https://doi.org/10.1103/PhysRevLett.108.070604} {\bibfield  {journal} {\bibinfo  {journal} {Phys. Rev. Lett.}\ }\textbf {\bibinfo {volume} {108}},\ \bibinfo {pages} {070604} (\bibinfo {year} {2012})}\BibitemShut {NoStop}%
\bibitem [{\citenamefont {Brunner}\ \emph {et~al.}(2014)\citenamefont {Brunner}, \citenamefont {Huber}, \citenamefont {Linden}, \citenamefont {Popescu}, \citenamefont {Silva},\ and\ \citenamefont {Skrzypczyk}}]{Brunner2014}%
  \BibitemOpen
  \bibfield  {author} {\bibinfo {author} {\bibfnamefont {N.}~\bibnamefont {Brunner}}, \bibinfo {author} {\bibfnamefont {M.}~\bibnamefont {Huber}}, \bibinfo {author} {\bibfnamefont {N.}~\bibnamefont {Linden}}, \bibinfo {author} {\bibfnamefont {S.}~\bibnamefont {Popescu}}, \bibinfo {author} {\bibfnamefont {R.}~\bibnamefont {Silva}},\ and\ \bibinfo {author} {\bibfnamefont {P.}~\bibnamefont {Skrzypczyk}},\ }\bibfield  {title} {\bibinfo {title} {Entanglement enhances cooling in microscopic quantum refrigerators},\ }\href {https://doi.org/10.1103/PhysRevE.89.032115} {\bibfield  {journal} {\bibinfo  {journal} {Phys. Rev. E}\ }\textbf {\bibinfo {volume} {89}},\ \bibinfo {pages} {032115} (\bibinfo {year} {2014})}\BibitemShut {NoStop}%
\bibitem [{\citenamefont {Correa}\ \emph {et~al.}(2014)\citenamefont {Correa}, \citenamefont {Palao}, \citenamefont {Alonso},\ and\ \citenamefont {Adesso}}]{Correa2014}%
  \BibitemOpen
  \bibfield  {author} {\bibinfo {author} {\bibfnamefont {L.~A.}\ \bibnamefont {Correa}}, \bibinfo {author} {\bibfnamefont {J.~P.}\ \bibnamefont {Palao}}, \bibinfo {author} {\bibfnamefont {D.}~\bibnamefont {Alonso}},\ and\ \bibinfo {author} {\bibfnamefont {G.}~\bibnamefont {Adesso}},\ }\bibfield  {title} {\bibinfo {title} {Quantum-enhanced absorption refrigerators},\ }\href {https://doi.org/10.1038/srep03949} {\bibfield  {journal} {\bibinfo  {journal} {Scientific Reports}\ }\textbf {\bibinfo {volume} {4}},\ \bibinfo {pages} {3949} (\bibinfo {year} {2014})}\BibitemShut {NoStop}%
\bibitem [{\citenamefont {Brask}\ and\ \citenamefont {Brunner}(2015)}]{Brask2015}%
  \BibitemOpen
  \bibfield  {author} {\bibinfo {author} {\bibfnamefont {J.~B.}\ \bibnamefont {Brask}}\ and\ \bibinfo {author} {\bibfnamefont {N.}~\bibnamefont {Brunner}},\ }\bibfield  {title} {\bibinfo {title} {Small quantum absorption refrigerator in the transient regime: Time scales, enhanced cooling, and entanglement},\ }\href {https://doi.org/10.1103/PhysRevE.92.062101} {\bibfield  {journal} {\bibinfo  {journal} {Phys. Rev. E}\ }\textbf {\bibinfo {volume} {92}},\ \bibinfo {pages} {062101} (\bibinfo {year} {2015})}\BibitemShut {NoStop}%
\bibitem [{\citenamefont {Wang}\ \emph {et~al.}(2015)\citenamefont {Wang}, \citenamefont {Lai}, \citenamefont {Ye}, \citenamefont {He}, \citenamefont {Ma},\ and\ \citenamefont {Liao}}]{Wang2015}%
  \BibitemOpen
  \bibfield  {author} {\bibinfo {author} {\bibfnamefont {J.}~\bibnamefont {Wang}}, \bibinfo {author} {\bibfnamefont {Y.}~\bibnamefont {Lai}}, \bibinfo {author} {\bibfnamefont {Z.}~\bibnamefont {Ye}}, \bibinfo {author} {\bibfnamefont {J.}~\bibnamefont {He}}, \bibinfo {author} {\bibfnamefont {Y.}~\bibnamefont {Ma}},\ and\ \bibinfo {author} {\bibfnamefont {Q.}~\bibnamefont {Liao}},\ }\bibfield  {title} {\bibinfo {title} {Four-level refrigerator driven by photons},\ }\href {https://doi.org/10.1103/PhysRevE.91.050102} {\bibfield  {journal} {\bibinfo  {journal} {Phys. Rev. E}\ }\textbf {\bibinfo {volume} {91}},\ \bibinfo {pages} {050102} (\bibinfo {year} {2015})}\BibitemShut {NoStop}%
\bibitem [{\citenamefont {Mitchison}\ \emph {et~al.}(2015)\citenamefont {Mitchison}, \citenamefont {Woods}, \citenamefont {Prior},\ and\ \citenamefont {Huber}}]{Mitchison2015}%
  \BibitemOpen
  \bibfield  {author} {\bibinfo {author} {\bibfnamefont {M.~T.}\ \bibnamefont {Mitchison}}, \bibinfo {author} {\bibfnamefont {M.~P.}\ \bibnamefont {Woods}}, \bibinfo {author} {\bibfnamefont {J.}~\bibnamefont {Prior}},\ and\ \bibinfo {author} {\bibfnamefont {M.}~\bibnamefont {Huber}},\ }\bibfield  {title} {\bibinfo {title} {Coherence-assisted single-shot cooling by quantum absorption refrigerators},\ }\href {https://doi.org/10.1088/1367-2630/17/11/115013} {\bibfield  {journal} {\bibinfo  {journal} {New J. Phys.}\ }\textbf {\bibinfo {volume} {17}},\ \bibinfo {pages} {115013} (\bibinfo {year} {2015})}\BibitemShut {NoStop}%
\bibitem [{\citenamefont {Mu}\ \emph {et~al.}(2017)\citenamefont {Mu}, \citenamefont {Agarwalla}, \citenamefont {Schaller},\ and\ \citenamefont {Segal}}]{Mu2017}%
  \BibitemOpen
  \bibfield  {author} {\bibinfo {author} {\bibfnamefont {A.}~\bibnamefont {Mu}}, \bibinfo {author} {\bibfnamefont {B.~K.}\ \bibnamefont {Agarwalla}}, \bibinfo {author} {\bibfnamefont {G.}~\bibnamefont {Schaller}},\ and\ \bibinfo {author} {\bibfnamefont {D.}~\bibnamefont {Segal}},\ }\bibfield  {title} {\bibinfo {title} {Qubit absorption refrigerator at strong coupling},\ }\href {https://doi.org/10.1088/1367-2630/aa9b75} {\bibfield  {journal} {\bibinfo  {journal} {New J. Phys.}\ }\textbf {\bibinfo {volume} {19}},\ \bibinfo {pages} {123034} (\bibinfo {year} {2017})}\BibitemShut {NoStop}%
\bibitem [{\citenamefont {Nimmrichter}\ \emph {et~al.}(2017)\citenamefont {Nimmrichter}, \citenamefont {Dai}, \citenamefont {Roulet},\ and\ \citenamefont {Scarani}}]{Nimmrichter2017}%
  \BibitemOpen
  \bibfield  {author} {\bibinfo {author} {\bibfnamefont {S.}~\bibnamefont {Nimmrichter}}, \bibinfo {author} {\bibfnamefont {J.}~\bibnamefont {Dai}}, \bibinfo {author} {\bibfnamefont {A.}~\bibnamefont {Roulet}},\ and\ \bibinfo {author} {\bibfnamefont {V.}~\bibnamefont {Scarani}},\ }\bibfield  {title} {\bibinfo {title} {Quantum and classical dynamics of a three-mode absorption refrigerator},\ }\href {https://doi.org/10.22331/q-2017-12-11-37} {\bibfield  {journal} {\bibinfo  {journal} {Quantum}\ }\textbf {\bibinfo {volume} {1}},\ \bibinfo {pages} {37} (\bibinfo {year} {2017})}\BibitemShut {NoStop}%
\bibitem [{\citenamefont {Mukhopadhyay}\ \emph {et~al.}(2018)\citenamefont {Mukhopadhyay}, \citenamefont {Misra}, \citenamefont {Bhattacharya},\ and\ \citenamefont {Pati}}]{Mukhopadhyay2018}%
  \BibitemOpen
  \bibfield  {author} {\bibinfo {author} {\bibfnamefont {C.}~\bibnamefont {Mukhopadhyay}}, \bibinfo {author} {\bibfnamefont {A.}~\bibnamefont {Misra}}, \bibinfo {author} {\bibfnamefont {S.}~\bibnamefont {Bhattacharya}},\ and\ \bibinfo {author} {\bibfnamefont {A.~K.}\ \bibnamefont {Pati}},\ }\bibfield  {title} {\bibinfo {title} {Quantum speed limit constraints on a nanoscale autonomous refrigerator},\ }\href {https://doi.org/10.1103/PhysRevE.97.062116} {\bibfield  {journal} {\bibinfo  {journal} {Phys. Rev. E}\ }\textbf {\bibinfo {volume} {97}},\ \bibinfo {pages} {062116} (\bibinfo {year} {2018})}\BibitemShut {NoStop}%
\bibitem [{\citenamefont {Mitchison}(2019)}]{Mitchison2019}%
  \BibitemOpen
  \bibfield  {author} {\bibinfo {author} {\bibfnamefont {M.~T.}\ \bibnamefont {Mitchison}},\ }\bibfield  {title} {\bibinfo {title} {Quantum thermal absorption machines: refrigerators, engines and clocks},\ }\href {https://doi.org/10.1080/00107514.2019.1631555} {\bibfield  {journal} {\bibinfo  {journal} {Contemporary Physics}\ }\textbf {\bibinfo {volume} {60}},\ \bibinfo {pages} {164–187} (\bibinfo {year} {2019})}\BibitemShut {NoStop}%
\bibitem [{\citenamefont {Das}\ \emph {et~al.}(2019)\citenamefont {Das}, \citenamefont {Misra}, \citenamefont {Pal}, \citenamefont {Sen(De)},\ and\ \citenamefont {Sen}}]{Das2019}%
  \BibitemOpen
  \bibfield  {author} {\bibinfo {author} {\bibfnamefont {S.}~\bibnamefont {Das}}, \bibinfo {author} {\bibfnamefont {A.}~\bibnamefont {Misra}}, \bibinfo {author} {\bibfnamefont {A.~K.}\ \bibnamefont {Pal}}, \bibinfo {author} {\bibfnamefont {A.}~\bibnamefont {Sen(De)}},\ and\ \bibinfo {author} {\bibfnamefont {U.}~\bibnamefont {Sen}},\ }\bibfield  {title} {\bibinfo {title} {Necessarily transient quantum refrigerator},\ }\href {https://doi.org/10.1209/0295-5075/125/20007} {\bibfield  {journal} {\bibinfo  {journal} {Europhysics Letters}\ }\textbf {\bibinfo {volume} {125}},\ \bibinfo {pages} {20007} (\bibinfo {year} {2019})}\BibitemShut {NoStop}%
\bibitem [{\citenamefont {Hewgill}\ \emph {et~al.}(2020)\citenamefont {Hewgill}, \citenamefont {Gonz\'alez}, \citenamefont {Palao}, \citenamefont {Alonso}, \citenamefont {Ferraro},\ and\ \citenamefont {De~Chiara}}]{Hewgill2020}%
  \BibitemOpen
  \bibfield  {author} {\bibinfo {author} {\bibfnamefont {A.}~\bibnamefont {Hewgill}}, \bibinfo {author} {\bibfnamefont {J.~O.}\ \bibnamefont {Gonz\'alez}}, \bibinfo {author} {\bibfnamefont {J.~P.}\ \bibnamefont {Palao}}, \bibinfo {author} {\bibfnamefont {D.}~\bibnamefont {Alonso}}, \bibinfo {author} {\bibfnamefont {A.}~\bibnamefont {Ferraro}},\ and\ \bibinfo {author} {\bibfnamefont {G.}~\bibnamefont {De~Chiara}},\ }\bibfield  {title} {\bibinfo {title} {Three-qubit refrigerator with two-body interactions},\ }\href {https://doi.org/10.1103/PhysRevE.101.012109} {\bibfield  {journal} {\bibinfo  {journal} {Phys. Rev. E}\ }\textbf {\bibinfo {volume} {101}},\ \bibinfo {pages} {012109} (\bibinfo {year} {2020})}\BibitemShut {NoStop}%
\bibitem [{\citenamefont {Ghoshal}\ \emph {et~al.}(2021)\citenamefont {Ghoshal}, \citenamefont {Das}, \citenamefont {Pal}, \citenamefont {Sen(De)},\ and\ \citenamefont {Sen}}]{Ghoshal2021}%
  \BibitemOpen
  \bibfield  {author} {\bibinfo {author} {\bibfnamefont {A.}~\bibnamefont {Ghoshal}}, \bibinfo {author} {\bibfnamefont {S.}~\bibnamefont {Das}}, \bibinfo {author} {\bibfnamefont {A.~K.}\ \bibnamefont {Pal}}, \bibinfo {author} {\bibfnamefont {A.}~\bibnamefont {Sen(De)}},\ and\ \bibinfo {author} {\bibfnamefont {U.}~\bibnamefont {Sen}},\ }\bibfield  {title} {\bibinfo {title} {Three qubits in less than three baths: Beyond two-body system-bath interactions in quantum refrigerators},\ }\href {https://doi.org/10.1103/PhysRevA.104.042208} {\bibfield  {journal} {\bibinfo  {journal} {Phys. Rev. A}\ }\textbf {\bibinfo {volume} {104}},\ \bibinfo {pages} {042208} (\bibinfo {year} {2021})}\BibitemShut {NoStop}%
\bibitem [{\citenamefont {Ray}\ \emph {et~al.}(2023)\citenamefont {Ray}, \citenamefont {Mondal}, \citenamefont {Bhattacharyya}, \citenamefont {Ghoshal}, \citenamefont {Rakshit},\ and\ \citenamefont {Sen}}]{Ray2023}%
  \BibitemOpen
  \bibfield  {author} {\bibinfo {author} {\bibfnamefont {T.}~\bibnamefont {Ray}}, \bibinfo {author} {\bibfnamefont {S.}~\bibnamefont {Mondal}}, \bibinfo {author} {\bibfnamefont {A.}~\bibnamefont {Bhattacharyya}}, \bibinfo {author} {\bibfnamefont {A.}~\bibnamefont {Ghoshal}}, \bibinfo {author} {\bibfnamefont {D.}~\bibnamefont {Rakshit}},\ and\ \bibinfo {author} {\bibfnamefont {U.}~\bibnamefont {Sen}},\ }\bibfield  {title} {\bibinfo {title} {Kerr-type nonlinear baths enhance cooling in quantum refrigerators},\ }\href {https://doi.org/10.48550/arXiv.2311.10499} {\bibfield  {journal} {\bibinfo  {journal} {arXiv:2311.10499}\ } (\bibinfo {year} {2023})}\BibitemShut {NoStop}%
\bibitem [{\citenamefont {Bhattacharyya}\ \emph {et~al.}(2025)\citenamefont {Bhattacharyya}, \citenamefont {Ghoshal},\ and\ \citenamefont {Sen}}]{Bhattacharyya2025}%
  \BibitemOpen
  \bibfield  {author} {\bibinfo {author} {\bibfnamefont {A.}~\bibnamefont {Bhattacharyya}}, \bibinfo {author} {\bibfnamefont {A.}~\bibnamefont {Ghoshal}},\ and\ \bibinfo {author} {\bibfnamefont {U.}~\bibnamefont {Sen}},\ }\bibfield  {title} {\bibinfo {title} {Transient effects in quantum refrigerators with finite environments},\ }\href {https://doi.org/10.1103/PhysRevA.111.012209} {\bibfield  {journal} {\bibinfo  {journal} {Phys. Rev. A}\ }\textbf {\bibinfo {volume} {111}},\ \bibinfo {pages} {012209} (\bibinfo {year} {2025})}\BibitemShut {NoStop}%
\bibitem [{\citenamefont {Mondkar}\ \emph {et~al.}(2025)\citenamefont {Mondkar}, \citenamefont {Bhattacharyya},\ and\ \citenamefont {Sen}}]{Mondkar2025}%
  \BibitemOpen
  \bibfield  {author} {\bibinfo {author} {\bibfnamefont {S.}~\bibnamefont {Mondkar}}, \bibinfo {author} {\bibfnamefont {A.}~\bibnamefont {Bhattacharyya}},\ and\ \bibinfo {author} {\bibfnamefont {U.}~\bibnamefont {Sen}},\ }\bibfield  {title} {\bibinfo {title} {Quantum refrigerator embedded in spin-star environments: Scalings of temperature and refrigeration time},\ }\href {https://doi.org/10.48550/arXiv.2505.04374} {\bibfield  {journal} {\bibinfo  {journal} {arXiv:2505.04374}\ } (\bibinfo {year} {2025})}\BibitemShut {NoStop}%
\bibitem [{\citenamefont {Huang}\ \emph {et~al.}(2024)\citenamefont {Huang}, \citenamefont {Xi}, \citenamefont {Long}, \citenamefont {Liu}, \citenamefont {Fan}, \citenamefont {Wang}, \citenamefont {Zheng}, \citenamefont {Feng}, \citenamefont {Nie},\ and\ \citenamefont {Lu}}]{Huang2024}%
  \BibitemOpen
  \bibfield  {author} {\bibinfo {author} {\bibfnamefont {K.}~\bibnamefont {Huang}}, \bibinfo {author} {\bibfnamefont {C.}~\bibnamefont {Xi}}, \bibinfo {author} {\bibfnamefont {X.}~\bibnamefont {Long}}, \bibinfo {author} {\bibfnamefont {H.}~\bibnamefont {Liu}}, \bibinfo {author} {\bibfnamefont {Y.-a.}\ \bibnamefont {Fan}}, \bibinfo {author} {\bibfnamefont {X.}~\bibnamefont {Wang}}, \bibinfo {author} {\bibfnamefont {Y.}~\bibnamefont {Zheng}}, \bibinfo {author} {\bibfnamefont {Y.}~\bibnamefont {Feng}}, \bibinfo {author} {\bibfnamefont {X.}~\bibnamefont {Nie}},\ and\ \bibinfo {author} {\bibfnamefont {D.}~\bibnamefont {Lu}},\ }\bibfield  {title} {\bibinfo {title} {Experimental realization of self-contained quantum refrigeration},\ }\href {https://doi.org/10.1103/PhysRevLett.132.210403} {\bibfield  {journal} {\bibinfo  {journal} {Phys. Rev. Lett.}\ }\textbf {\bibinfo {volume} {132}},\ \bibinfo {pages} {210403} (\bibinfo {year} {2024})}\BibitemShut {NoStop}%
\bibitem [{\citenamefont {Maslennikov}\ \emph {et~al.}(2019)\citenamefont {Maslennikov}, \citenamefont {Ding}, \citenamefont {Habl\"{u}tzel}, \citenamefont {Gan}, \citenamefont {Roulet}, \citenamefont {Nimmrichter}, \citenamefont {Dai}, \citenamefont {Scarani},\ and\ \citenamefont {Matsukevich}}]{Maslennikov2019}%
  \BibitemOpen
  \bibfield  {author} {\bibinfo {author} {\bibfnamefont {G.}~\bibnamefont {Maslennikov}}, \bibinfo {author} {\bibfnamefont {S.}~\bibnamefont {Ding}}, \bibinfo {author} {\bibfnamefont {R.}~\bibnamefont {Habl\"{u}tzel}}, \bibinfo {author} {\bibfnamefont {J.}~\bibnamefont {Gan}}, \bibinfo {author} {\bibfnamefont {A.}~\bibnamefont {Roulet}}, \bibinfo {author} {\bibfnamefont {S.}~\bibnamefont {Nimmrichter}}, \bibinfo {author} {\bibfnamefont {J.}~\bibnamefont {Dai}}, \bibinfo {author} {\bibfnamefont {V.}~\bibnamefont {Scarani}},\ and\ \bibinfo {author} {\bibfnamefont {D.}~\bibnamefont {Matsukevich}},\ }\bibfield  {title} {\bibinfo {title} {Quantum absorption refrigerator with trapped ions},\ }\href {http://dx.doi.org/10.1038/s41467-018-08090-0} {\bibfield  {journal} {\bibinfo  {journal} {Nat Commun}\ }\textbf {\bibinfo {volume} {10}} (\bibinfo {year} {2019})}\BibitemShut {NoStop}%
\bibitem [{\citenamefont {Yan}\ and\ \citenamefont {Jing}(2021)}]{Yan2021}%
  \BibitemOpen
  \bibfield  {author} {\bibinfo {author} {\bibfnamefont {J.-s.}\ \bibnamefont {Yan}}\ and\ \bibinfo {author} {\bibfnamefont {J.}~\bibnamefont {Jing}},\ }\bibfield  {title} {\bibinfo {title} {External-level assisted cooling by measurement},\ }\href {https://doi.org/10.1103/PhysRevA.104.063105} {\bibfield  {journal} {\bibinfo  {journal} {Phys. Rev. A}\ }\textbf {\bibinfo {volume} {104}},\ \bibinfo {pages} {063105} (\bibinfo {year} {2021})}\BibitemShut {NoStop}%
\bibitem [{\citenamefont {Yan}\ and\ \citenamefont {Jing}(2022)}]{Yan2022}%
  \BibitemOpen
  \bibfield  {author} {\bibinfo {author} {\bibfnamefont {J.-s.}\ \bibnamefont {Yan}}\ and\ \bibinfo {author} {\bibfnamefont {J.}~\bibnamefont {Jing}},\ }\bibfield  {title} {\bibinfo {title} {Simultaneous cooling by measuring one ancillary system},\ }\href {https://doi.org/10.1103/PhysRevA.105.052607} {\bibfield  {journal} {\bibinfo  {journal} {Phys. Rev. A}\ }\textbf {\bibinfo {volume} {105}},\ \bibinfo {pages} {052607} (\bibinfo {year} {2022})}\BibitemShut {NoStop}%
\bibitem [{\citenamefont {Konar}\ \emph {et~al.}(2022)\citenamefont {Konar}, \citenamefont {Ghosh},\ and\ \citenamefont {Sen(De)}}]{Konar2022}%
  \BibitemOpen
  \bibfield  {author} {\bibinfo {author} {\bibfnamefont {T.~K.}\ \bibnamefont {Konar}}, \bibinfo {author} {\bibfnamefont {S.}~\bibnamefont {Ghosh}},\ and\ \bibinfo {author} {\bibfnamefont {A.}~\bibnamefont {Sen(De)}},\ }\bibfield  {title} {\bibinfo {title} {Refrigeration via purification through repeated measurements},\ }\href {https://doi.org/10.1103/PhysRevA.106.022616} {\bibfield  {journal} {\bibinfo  {journal} {Phys. Rev. A}\ }\textbf {\bibinfo {volume} {106}},\ \bibinfo {pages} {022616} (\bibinfo {year} {2022})}\BibitemShut {NoStop}%
\bibitem [{\citenamefont {Ghosh}\ \emph {et~al.}(2024)\citenamefont {Ghosh}, \citenamefont {Konar},\ and\ \citenamefont {Sen(De)}}]{Ghosh2024}%
  \BibitemOpen
  \bibfield  {author} {\bibinfo {author} {\bibfnamefont {D.}~\bibnamefont {Ghosh}}, \bibinfo {author} {\bibfnamefont {T.~K.}\ \bibnamefont {Konar}},\ and\ \bibinfo {author} {\bibfnamefont {A.}~\bibnamefont {Sen(De)}},\ }\bibfield  {title} {\bibinfo {title} {Measurement-based qudit quantum refrigerator with subspace cooling},\ }\href {https://doi.org/10.48550/arXiv.2409.08375} {\bibfield  {journal} {\bibinfo  {journal} {arXiv:2409.08375}\ } (\bibinfo {year} {2024})}\BibitemShut {NoStop}%
\bibitem [{\citenamefont {Brunner}\ \emph {et~al.}(2012)\citenamefont {Brunner}, \citenamefont {Linden}, \citenamefont {Popescu},\ and\ \citenamefont {Skrzypczyk}}]{Brunner2012}%
  \BibitemOpen
  \bibfield  {author} {\bibinfo {author} {\bibfnamefont {N.}~\bibnamefont {Brunner}}, \bibinfo {author} {\bibfnamefont {N.}~\bibnamefont {Linden}}, \bibinfo {author} {\bibfnamefont {S.}~\bibnamefont {Popescu}},\ and\ \bibinfo {author} {\bibfnamefont {P.}~\bibnamefont {Skrzypczyk}},\ }\bibfield  {title} {\bibinfo {title} {Virtual qubits, virtual temperatures, and the foundations of thermodynamics},\ }\href {https://doi.org/10.1103/PhysRevE.85.051117} {\bibfield  {journal} {\bibinfo  {journal} {Phys. Rev. E}\ }\textbf {\bibinfo {volume} {85}},\ \bibinfo {pages} {051117} (\bibinfo {year} {2012})}\BibitemShut {NoStop}%
\bibitem [{\citenamefont {Ghanavati}\ and\ \citenamefont {Movahhedian}(2014)}]{Ghanavati2014}%
  \BibitemOpen
  \bibfield  {author} {\bibinfo {author} {\bibfnamefont {M.}~\bibnamefont {Ghanavati}}\ and\ \bibinfo {author} {\bibfnamefont {H.}~\bibnamefont {Movahhedian}},\ }\bibfield  {title} {\bibinfo {title} {Self-contained n-qubit quantum refrigerator},\ }\href {https://doi.org/10.1142/S021974991450018X} {\bibfield  {journal} {\bibinfo  {journal} {International Journal of Quantum Information}\ }\textbf {\bibinfo {volume} {12}},\ \bibinfo {pages} {1450018} (\bibinfo {year} {2014})}\BibitemShut {NoStop}%
\bibitem [{\citenamefont {Konar}\ \emph {et~al.}(2023)\citenamefont {Konar}, \citenamefont {Ghosh}, \citenamefont {Pal},\ and\ \citenamefont {Sen(De)}}]{Konar2023}%
  \BibitemOpen
  \bibfield  {author} {\bibinfo {author} {\bibfnamefont {T.~K.}\ \bibnamefont {Konar}}, \bibinfo {author} {\bibfnamefont {S.}~\bibnamefont {Ghosh}}, \bibinfo {author} {\bibfnamefont {A.~K.}\ \bibnamefont {Pal}},\ and\ \bibinfo {author} {\bibfnamefont {A.}~\bibnamefont {Sen(De)}},\ }\bibfield  {title} {\bibinfo {title} {Designing refrigerators in higher dimensions using quantum spin models},\ }\href {https://link.aps.org/doi/10.1103/PhysRevA.107.032602} {\bibfield  {journal} {\bibinfo  {journal} {Phys. Rev. A}\ }\textbf {\bibinfo {volume} {107}},\ \bibinfo {pages} {032602} (\bibinfo {year} {2023})}\BibitemShut {NoStop}%
\bibitem [{\citenamefont {Krishnan}\ \emph {et~al.}(2024)\citenamefont {Krishnan}, \citenamefont {Pushpan},\ and\ \citenamefont {Pal}}]{Krishnan2024}%
  \BibitemOpen
  \bibfield  {author} {\bibinfo {author} {\bibfnamefont {J.~G.}\ \bibnamefont {Krishnan}}, \bibinfo {author} {\bibfnamefont {C.~B.}\ \bibnamefont {Pushpan}},\ and\ \bibinfo {author} {\bibfnamefont {A.~K.}\ \bibnamefont {Pal}},\ }\bibfield  {title} {\bibinfo {title} {Simultaneous cooling of qubits via a quantum absorption refrigerator and beyond},\ }\href {https://doi.org/10.48550/arXiv.2410.15871} {\bibfield  {journal} {\bibinfo  {journal} {arXiv:2410.15871}\ } (\bibinfo {year} {2024})}\BibitemShut {NoStop}%
\bibitem [{\citenamefont {Guo}\ \emph {et~al.}(2018)\citenamefont {Guo}, \citenamefont {Liu},\ and\ \citenamefont {Yu}}]{Guo2018}%
  \BibitemOpen
  \bibfield  {author} {\bibinfo {author} {\bibfnamefont {B.-q.}\ \bibnamefont {Guo}}, \bibinfo {author} {\bibfnamefont {T.}~\bibnamefont {Liu}},\ and\ \bibinfo {author} {\bibfnamefont {C.-s.}\ \bibnamefont {Yu}},\ }\bibfield  {title} {\bibinfo {title} {Quantum thermal transistor based on qubit-qutrit coupling},\ }\href {https://doi.org/10.1103/PhysRevE.98.022118} {\bibfield  {journal} {\bibinfo  {journal} {Phys. Rev. E}\ }\textbf {\bibinfo {volume} {98}},\ \bibinfo {pages} {022118} (\bibinfo {year} {2018})}\BibitemShut {NoStop}%
\bibitem [{\citenamefont {Breuer}\ and\ \citenamefont {Petruccione}(2002)}]{Petruccione2002}%
  \BibitemOpen
  \bibfield  {author} {\bibinfo {author} {\bibfnamefont {H.~P.}\ \bibnamefont {Breuer}}\ and\ \bibinfo {author} {\bibfnamefont {F.}~\bibnamefont {Petruccione}},\ }\href@noop {} {\emph {\bibinfo {title} {The theory of open quantum systems}}}\ (\bibinfo  {publisher} {Oxford University Press},\ \bibinfo {address} {Great Clarendon Street},\ \bibinfo {year} {2002})\BibitemShut {NoStop}%
\bibitem [{\citenamefont {Nielsen}\ and\ \citenamefont {Chuang}(2011)}]{Nielsen2011}%
  \BibitemOpen
  \bibfield  {author} {\bibinfo {author} {\bibfnamefont {M.~A.}\ \bibnamefont {Nielsen}}\ and\ \bibinfo {author} {\bibfnamefont {I.~L.}\ \bibnamefont {Chuang}},\ }\href@noop {} {\emph {\bibinfo {title} {Quantum Computation and Quantum Information: 10th Anniversary Edition}}}\ (\bibinfo  {publisher} {Cambridge University Press},\ \bibinfo {year} {2011})\BibitemShut {NoStop}%
\bibitem [{\citenamefont {Rivas}\ and\ \citenamefont {Huelga}(2012)}]{Rivas2012}%
  \BibitemOpen
  \bibfield  {author} {\bibinfo {author} {\bibfnamefont {A.}~\bibnamefont {Rivas}}\ and\ \bibinfo {author} {\bibfnamefont {S.~F.}\ \bibnamefont {Huelga}},\ }\href@noop {} {\emph {\bibinfo {title} {Open quantum systems}}},\ Vol.~\bibinfo {volume} {10}\ (\bibinfo  {publisher} {Springer},\ \bibinfo {year} {2012})\BibitemShut {NoStop}%
\bibitem [{\citenamefont {Medina}\ \emph {et~al.}(2025)\citenamefont {Medina}, \citenamefont {Culhane}, \citenamefont {Binder}, \citenamefont {Landi},\ and\ \citenamefont {Goold}}]{Medina2025}%
  \BibitemOpen
  \bibfield  {author} {\bibinfo {author} {\bibfnamefont {I.}~\bibnamefont {Medina}}, \bibinfo {author} {\bibfnamefont {O.}~\bibnamefont {Culhane}}, \bibinfo {author} {\bibfnamefont {F.~C.}\ \bibnamefont {Binder}}, \bibinfo {author} {\bibfnamefont {G.~T.}\ \bibnamefont {Landi}},\ and\ \bibinfo {author} {\bibfnamefont {J.}~\bibnamefont {Goold}},\ }\bibfield  {title} {\bibinfo {title} {Anomalous discharging of quantum batteries: The ergotropic mpemba effect},\ }\href {https://doi.org/10.1103/PhysRevLett.134.220402} {\bibfield  {journal} {\bibinfo  {journal} {Phys. Rev. Lett.}\ }\textbf {\bibinfo {volume} {134}},\ \bibinfo {pages} {220402} (\bibinfo {year} {2025})}\BibitemShut {NoStop}%
\bibitem [{\citenamefont {Liu}\ \emph {et~al.}(2024{\natexlab{c}})\citenamefont {Liu}, \citenamefont {Yuan}, \citenamefont {Ruan}, \citenamefont {Xu}, \citenamefont {Luo}, \citenamefont {He}, \citenamefont {He}, \citenamefont {Ma},\ and\ \citenamefont {Wang}}]{Liu2024}%
  \BibitemOpen
  \bibfield  {author} {\bibinfo {author} {\bibfnamefont {D.}~\bibnamefont {Liu}}, \bibinfo {author} {\bibfnamefont {J.}~\bibnamefont {Yuan}}, \bibinfo {author} {\bibfnamefont {H.}~\bibnamefont {Ruan}}, \bibinfo {author} {\bibfnamefont {Y.}~\bibnamefont {Xu}}, \bibinfo {author} {\bibfnamefont {S.}~\bibnamefont {Luo}}, \bibinfo {author} {\bibfnamefont {J.}~\bibnamefont {He}}, \bibinfo {author} {\bibfnamefont {X.}~\bibnamefont {He}}, \bibinfo {author} {\bibfnamefont {Y.}~\bibnamefont {Ma}},\ and\ \bibinfo {author} {\bibfnamefont {J.}~\bibnamefont {Wang}},\ }\bibfield  {title} {\bibinfo {title} {Speeding up quantum heat engines by the mpemba effect},\ }\href {https://doi.org/10.1103/PhysRevA.110.042218} {\bibfield  {journal} {\bibinfo  {journal} {Phys. Rev. A}\ }\textbf {\bibinfo {volume} {110}},\ \bibinfo {pages} {042218} (\bibinfo {year} {2024}{\natexlab{c}})}\BibitemShut {NoStop}%
\bibitem [{\citenamefont {Edo}\ and\ \citenamefont {Wu}(2024)}]{Edo2024}%
  \BibitemOpen
  \bibfield  {author} {\bibinfo {author} {\bibfnamefont {M.~E.}\ \bibnamefont {Edo}}\ and\ \bibinfo {author} {\bibfnamefont {L.-A.}\ \bibnamefont {Wu}},\ }\bibfield  {title} {\bibinfo {title} {Study on quantum thermalization from thermal initial states in a superconducting quantum computer},\ }\href {https://doi.org/10.48550/arXiv.2403.14630} {\bibfield  {journal} {\bibinfo  {journal} {arXiv:2403.14630}\ } (\bibinfo {year} {2024})}\BibitemShut {NoStop}%
\bibitem [{\citenamefont {Campaioli}\ \emph {et~al.}(2024)\citenamefont {Campaioli}, \citenamefont {Cole},\ and\ \citenamefont {Hapuarachchi}}]{Campaioli2024}%
  \BibitemOpen
  \bibfield  {author} {\bibinfo {author} {\bibfnamefont {F.}~\bibnamefont {Campaioli}}, \bibinfo {author} {\bibfnamefont {J.~H.}\ \bibnamefont {Cole}},\ and\ \bibinfo {author} {\bibfnamefont {H.}~\bibnamefont {Hapuarachchi}},\ }\bibfield  {title} {\bibinfo {title} {Quantum master equations: Tips and tricks for quantum optics, quantum computing, and beyond},\ }\href {https://doi.org/10.1103/PRXQuantum.5.020202} {\bibfield  {journal} {\bibinfo  {journal} {PRX Quantum}\ }\textbf {\bibinfo {volume} {5}},\ \bibinfo {pages} {020202} (\bibinfo {year} {2024})}\BibitemShut {NoStop}%
\bibitem [{\citenamefont {Bhatia}(1997)}]{Bhatia-book}%
  \BibitemOpen
  \bibfield  {author} {\bibinfo {author} {\bibfnamefont {R.}~\bibnamefont {Bhatia}},\ }\href {https://doi.org/10.1007/978-1-4612-0653-8} {\emph {\bibinfo {title} {Matrix Analysis}}}\ (\bibinfo  {publisher} {Springer New York},\ \bibinfo {year} {1997})\BibitemShut {NoStop}%
\bibitem [{\citenamefont {Van~Vu}\ and\ \citenamefont {Hayakawa}(2025)}]{VanVu2025}%
  \BibitemOpen
  \bibfield  {author} {\bibinfo {author} {\bibfnamefont {T.}~\bibnamefont {Van~Vu}}\ and\ \bibinfo {author} {\bibfnamefont {H.}~\bibnamefont {Hayakawa}},\ }\bibfield  {title} {\bibinfo {title} {Thermomajorization mpemba effect},\ }\href {https://doi.org/10.1103/PhysRevLett.134.107101} {\bibfield  {journal} {\bibinfo  {journal} {Phys. Rev. Lett.}\ }\textbf {\bibinfo {volume} {134}},\ \bibinfo {pages} {107101} (\bibinfo {year} {2025})}\BibitemShut {NoStop}%
\bibitem [{\citenamefont {Johnson}(2007)}]{NLopt}%
  \BibitemOpen
  \bibfield  {author} {\bibinfo {author} {\bibfnamefont {S.~G.}\ \bibnamefont {Johnson}},\ }\href@noop {} {\bibinfo {title} {The {NLopt} nonlinear-optimization package}},\ \bibinfo {howpublished} {\url{https://github.com/stevengj/nlopt}} (\bibinfo {year} {2007})\BibitemShut {NoStop}%
\bibitem [{mon()}]{mondal_2025_17759279}%
  \BibitemOpen
  \bibfield  {title} {\bibinfo {title} {Dataset of ``{M}pemba effect in self-contained quantum refrigerators: accelerated cooling"},\ }\href {https://doi.org/10.5281/zenodo.17759279} {10.5281/zenodo.17759279}\BibitemShut {NoStop}%
\end{thebibliography}%

\end{document}